\documentclass{article}
\usepackage{graphicx} 

\title{Finer-Grained Hardness of Kernel Density Estimation}
\author{Josh Alman\footnote{Columbia University. \texttt{josh@cs.columbia.edu}} \and Yunfeng Guan\footnote{Columbia University. \texttt{yunfeng.guan@columbia.edu}}} 

\usepackage{latexsym}
\usepackage{amsmath}
\usepackage{amssymb}
\usepackage{amsthm}
\usepackage{qcircuit}
\usepackage{bbm}

\usepackage{enumerate}
\usepackage{algorithm, algpseudocode}
\usepackage{tikz}
\usepackage{verbatim}
\usepackage{pict2e, xfp}
\usepackage{hyperref}

\newcommand{\handout}[5]{
  \noindent
  \begin{center}
  \framebox{
    \vbox{
      \hbox to 5.78in { #1 \hfill #2 }
      \vspace{4mm}
      \hbox to 5.78in { {\Large \hfill #5  \hfill} }
      \vspace{2mm}
      \hbox to 5.78in { {#3 \hfill #4} }
    }
  }
  \end{center}
  \vspace*{4mm}
}

\newtheorem{theorem}{Theorem}
\newtheorem{corollary}[theorem]{Corollary}
\newtheorem{lemma}[theorem]{Lemma}

\newtheorem{proposition}[theorem]{Proposition}
\newtheorem{definition}[theorem]{Definition}

\newtheorem{problem}[theorem]{Problem}
\newtheorem{conjecture}[theorem]{Conjecture}
\newtheorem{question}[theorem]{Question}

\theoremstyle{remark}
\newtheorem*{remark}{Remark}

\numberwithin{equation}{section}
\numberwithin{theorem}{section}

\newcommand{\ip}[2]{\langle #1 , #2 \rangle}

\DeclareMathOperator*{\Ex}{\mathbb{E}}
\newcommand{\poly}{\mathrm{poly}}

\newcommand{\N}{\ensuremath{\mathbb{N}}}
\newcommand{\F}{\ensuremath{\mathbb{F}}}

\newcommand{\R}{\ensuremath{\mathbb{R}}}
\newcommand{\Z}{\ensuremath{\mathbb{Z}}}

\newcommand{\eps}{\varepsilon}

\newcommand{\ind}{\mathbbm 1}
\newcommand{\bo}{\boldsymbol 1}

\newcommand{\KDE}{\mathsf{KDE}}
\newcommand{\bn}{\bar {n}}

\advance \topmargin by -\headheight
\advance \topmargin by -\headsep
\evensidemargin \oddsidemargin
\begin{document}
\maketitle

\begin{abstract}
    In batch Kernel Density Estimation (KDE) for a kernel function $f : \mathbb{R}^m \times \mathbb{R}^m \to \mathbb{R}$, we are given as input $2n$ points $x^{(1)}, \ldots, x^{(n)}, y^{(1)}, \ldots, y^{(n)} \in \mathbb{R}^m$ in dimension $m$, as well as a vector $v \in \mathbb{R}^n$. These inputs implicitly define the $n \times n$ kernel matrix $K$ given by $K[i,j] = f(x^{(i)}, y^{(j)})$. The goal is to compute a vector $v \in \mathbb{R}^n$ which approximates $K w$, i.e., with $|| Kw - v||_\infty < \varepsilon ||w||_1$.
    For illustrative purposes, consider the Gaussian kernel $f(x,y) := e^{-||x-y||_2^2}$. The classic approach to this problem is the famous Fast Multipole Method (FMM), which runs in time $n \cdot O(\log^m(\eps^{-1}))$ and is particularly effective in low dimensions because of its exponential dependence on $m$. Recently, as the higher-dimensional case $m \geq \Omega(\log n)$ has seen more applications in machine learning and statistics, new algorithms have focused on this setting: an algorithm using discrepancy theory, which runs in time $O(n / \eps)$, and an algorithm based on the polynomial method, which achieves inverse polynomial accuracy in almost linear time when the input points have bounded square diameter $B < o(\log n)$.
    A recent line of work has proved fine-grained lower bounds, with the goal of showing that the `curse of dimensionality' arising in FMM is necessary assuming the Strong Exponential Time Hypothesis (SETH). 
    Backurs et al. [NeurIPS 2017] first showed the hardness of a variety of Empirical Risk Minimization problems including KDE for Gaussian-like kernels in the case with high dimension $m = \Omega(\log n)$ and large scale $B = \Omega(\log n)$. Alman et al. [FOCS 2020] later developed new reductions in roughly this same parameter regime, leading to lower bounds for more general kernels, but only for very small error $\varepsilon < 2^{- \log^{\Omega(1)} (n)}$.

    In this paper, we refine the approach of Alman et al. to show new lower bounds in all parameter regimes, closing gaps between the known algorithms and lower bounds. For example:
    \begin{itemize}
        \item In the setting where $m = C\log n$ and $B = o(\log n)$, we prove Gaussian KDE requires $n^{2-o(1)}$ time to achieve additive error $\eps < \Omega(m/B)^{-m}$,  matching the performance of the polynomial method up to low-order terms.
        \item In the low dimensional setting $m = o(\log n)$, we show that Gaussian KDE requires $n^{2-o(1)}$ time to achieve $\eps$ such that $\log \log (\eps^{-1}) > \tilde \Omega ((\log n)/m)$, matching the error bound achievable by FMM up to low-order terms. To our knowledge, no nontrivial lower bound was previously known in this regime.
    \end{itemize} Our approach also generalizes to any parameter regime and any kernel. For example, we achieve similar fine-grained hardness results for any kernel with slowly-decaying Taylor coefficients such as the Cauchy kernel.
    Our new lower bounds make use of an intricate analysis of the `counting matrix', a special case of the kernel matrix focused on carefully-chosen evaluation points. As a key technical lemma, we give a novel approach to bounding the entries of its inverse by using Schur polynomials from algebraic combinatorics.
\end{abstract}

\thispagestyle{empty}
\newpage
\setcounter{page}{1}

\section{Introduction}
In computational statistics and learning theory, many applications reduce to solving the following problem: Given a set of points $X \subseteq \R^m$ sampled from some unknown distribution $\mathcal D$, estimate the probability density at a query point $y \in \R^m$ (or multiple such points $Y \subseteq \R^m$). This problem is known as the density estimation problem and has attracted interest in theoretical computer science in recent years. One of the most common methods for this problem is Kernel Density Estimation (KDE), which tries to approximate the distribution by a sum of kernel functions centered at each data point $x$. More concretely, if a kernel function $k: \R^m \times \R^m \rightarrow [0, 1]$ is appropriately picked, then the \emph{Kernel Density} $\mathrm{KDF}(y) := \frac 1n \sum_{x \in X} k(x, y)$ is a reasonably good approximation of $\mathcal D$ at point $y$.\footnote{For example, when $m = 1$ and $k(x, y) = \ind[|x - y| \le 1]$, the KDF is the histogram of the dataset $X$.}
In this work, we will focus on the popular class of radial kernels, i.e., kernels of the form $K(x, y) = f(\|x - y\|_2^2)$ for some function $f: \R_{\ge 0} \rightarrow [0, 1]$. Some prominent radial kernels include the Gaussian kernel with $f(u) = e^{-u}$, Rational Quadratic kernel with $f(u) = 1/(1+u)^{\sigma}$ for constant $\sigma$, and $t$-Student kernel with $f(u) = 1/(1+u^\rho)$ for constant $\rho$.

By virtue of its excellent statistical properties, Kernel Density Estimation has found numerous applications in computational statistics for tasks like mean estimation, classification, and outlier detection; see, for instance, the surveys~\cite{muandet2017kernel,chen2017tutorial} and~\cite[Chapter 1]{Sim19}. The prevalence of kernel methods in machine learning has also led to many new applications of Kernel Density Estimation~\cite{scholkopf2002learning}. One popular recent example is in `attention computation', the time bottleneck in computations involving transformers and other large language models; this is known to be essentially equivalent to Kernel Density Estimation with the Gaussian kernel~\cite{alman2023fast}\footnote{The parameters of KDE correspond directly to the parameters in attention: $n$ is the number of `tokens' in the query sequence, and $m$ is the dimension of the vectors which encode tokens.}.

In light of its wide applicability, tremendous effort has also been put into designing efficient algorithms for the KDE problem. The straightforward algorithm shows that for a given $y$, $\mathrm{KDF}(y)$ can be computed in $O(n)$ time, assuming $f$ is efficiently computable.\footnote{In this work we always assume $f(\|x-y\|_2^2)$ can be exactly computed in $O(1)$ time. Most of the previous works adopt this assumption.} If one aims at an exact result, this simple algorithm seems to be optimal. However, a number of advances starting from the celebrated Fast Multipole Method \cite{GR87} have successfully brought the running time down to $n^{o(1)}$ in various parameter regimes (after preprocessing the set $X$), provided an approximation to the Kernel Density suffices.
Before introducing these algorithmic ideas, we first give a formal definition of the (approximate) KDE problem. For simplicity, we present here the batched version, which asks to compute the KDF value for a collection of query points simultaneously. (We also compare the batched version and the data structure version in Section \ref{sec:discussion}.)
\begin{definition}[(Approximate) Kernel Density Estimation $\mathsf{KDE}_f(n, m, \eps, B)$]
Let $f: [0, B] \rightarrow [0, 1]$ be a real function and define the (kernel) function $k(x, y) = f(\|x-y\|_2^2)$.

Suppose we are given as input $2n$ points $x^{(1)}, \cdots, x^{(n)}, y^{(1)}, \cdots, y^{(n)} \in \R^m$ with the guarantee that $\|x^{(i)} - y^{(j)}\|_2^2 \le B$ for all $i, j \in [n]$.
Define the kernel matrix $K \in [0,1]^{n \times n}$ by $K[i, j] = f(\|x^{(i)}- y^{(j)}\|_2^2)$ for $i, j \in [n]$.
Then given additionally a vector $u \in \R^n$, one needs to output a vector $v \in \R^n$ such that
$\|v - K u\|_{\infty} \le \eps\|u\|_{1}$.
\end{definition}

\subsection{Algorithms}

As a computational problem, KDE has its running time dependent on four parameters: the number of data points / query points $n$, the dimension of vectors $m$, the \emph{additive} error of approximation $\eps$, and the maximum pairwise (square) distance $B$. Multiple parameter regimes arise from the their interplay, and the KDE problem tends to have rather different behavior across the regimes. Of course, the choice of the kernel function can also substantially change the complexity of the problem; in the following discussion we take the Gaussian kernel as a running example.

\paragraph{Low dimensional KDE: $m = o(\log n)$} The first nontrivial algorithm for the KDE problem is Greengard and Roklin's Fast Multipole Method \cite{GR87}. In this method, one partitions the space into bounded regions, and then Taylor expands the kernel around centers of these regions. In this way a truncation of the Taylor series yields a approximation of high accuracy when the data points lie in distant cells.
For the Gaussian kernel, the Fast Multipole Method runs in time $O(n\log^{O(m)}(n/\eps))$, which is exceptionally good in low dimensions (e.g., $m = O(1)$ in the original physical context\footnote{The Fast Multipole Method was originally introduced to solve the $n$-body problem from physics.} of \cite{GS91}).
However, in higher dimensional regimes, this approach suffers from an exponential dependence on $m$,
which is inherent in the (deterministic) space partitioning procedure (essentially building a quad-tree), and shared by other tree-based methods.

\paragraph{Moderate dimensional KDE: $m = \Theta(\log n)$} One way to avoid the exponential dependence on $m$ of the Fast Multipole Method is to use the method of polynomial approximation directly, without combining it with space partitioning. Alman and Aggarwal \cite{AA22} pinned down the optimal degree of a polynomial that approximates $f(u) = e^{-Bu}$ over $u \in [0, 1]$ by analyzing its Chebyshev truncation. A polynomial approximation of $e^{-Bu}$ then allows one to approximate the Gaussian kernel matrix using a matrix consisting of only polynomial entries. Such a matrix admits a decomposition as a product of two matrices of dimension $n \times n^{o(1)}$, for which the matrix-vector product can be performed efficiently.
As a result, they gave an algorithm for Gaussian KDE running in $n^{1+o(1)}$ time when $m = O(\log n), B = o(\log n)$ and $\eps = 1/\poly(n)$.
\paragraph{High dimensional KDE: $m = \omega(\log n)$} Another technique often used in the study of KDE is random sampling. For example, to compute $\frac 1n \sum_{x \in X} k(x, y)$, one can sample a subset $S \subseteq X$ so that $\frac 1{|S|} \sum_{x \in S} k(x, y)$ is a close approximation to $\mathrm{KDF}(y)$ when $|S|$ is sufficiently large. Simple calculation shows that $|S| = O(\log n/\eps^2)$ suffices, and a $\tilde O(n/\eps^2)$ time (randomized) algorithm follows. We note that this algorithm has no dependence on $m$ and works for arbitrarily high dimensions. This folklore random sampling algorithm stood unchallenged until recently Phillips and Tai \cite{PT20} devised an $O(n/\eps)$-time algorithm by building a small coreset based on discrepancy theory. 
They show that a clever subsampling scheme yields a smaller $S \subseteq X$ which has the same accuracy as its counterpart when used in the random sampling algorithm.
We in addition note that much effort \cite{CS17, CS19, BCIS18, BIW19, CKNS20, CKW24} has been dedicated to the \emph{relative} error setting, combining sampling schemes with techniques from high-dimensional geometry (such as hashing-based space partitioning). See Section \ref{sec:discussion} below where we compare the additive error and relative error settings and survey these algorithms in more detail.

\subsection{Lower bounds}
\label{sec:prevlb}
Despite the great variety of algorithms developed over the years, we still lack a comprehensive understanding of the complexity of the KDE problem. In this work, our goal is to complement the distinct algorithms targeting different parameter regimes with (nearly) matching running time lower bounds, and explain how the complexity of KDE is affected by parameters.

The first and most influential known lower bound was developed by Backurs, Indyk and Schimidt \cite{BIS17}. This work establishes a reduction from the Bichromatic Closest Pair (BCP) problem to a collection of empirical risk minimization problems including KDE. As BCP is a standard hard problem in fine-grained complexity assuming the Strong Exponential Time Hypothesis (SETH), the reduction leads to conditional lower bounds on the running time of KDE. To explain in detail, we first give the formal statement of the BCP problem.
\begin{problem}[Bichromatic Closest Pair]
$\emph{\textsf{Hamming-BCP}}(n, m)$: Given two sets $X = \{x^{(1)}, \cdots, x^{(n)}\}, Y = \{y^{(1)}, \cdots, y^{(n)}\} \in \{0, 1\}^m$, compute $\min_{i, j \in [n]} \|x^{(i)} - y^{(j)}\|_2^2$.
\end{problem}
We then observe that terms in the KDE result $\|K \times \bo\|_1$ can be grouped according to the pairwise distance between $x^{(i)}$ and $y^{(j)}$.
\[\|K \times \bo\|_1 = \sum_{i=1}^n\sum_{j=1}^n f(\|x^{(i)} - y^{(j)}\|_2^2)
= \sum_{p=0}^m f(p) \cdot \#\left\{(i, j)\in [n]^2: \|x^{(i)} - y^{(j)}\|_2^2 = p\right\}.\]
This identity provides a way of extracting minimum pairwise distance from the KDE result.
Indeed, if $\min_{i,j}\|x^{(i)} - y^{(j)}\|^2 \ge p+1$, then $\|K \times 1\|_1 \le n^2f(p+1)$ (assuming $f$ is decreasing); otherwise, there exists a pair $(x^{(i)}, y^{(j)})$ with $\|x^{(i)} - y^{(j)}\|^2 \le p$ and $\|K \times 1\|_1 \ge f(p)$. In this way, if $f$ is decreasing quickly enough, one can decide whether $\min_{i, j}\|x^{(i)} - y^{(j)}\|^2 \le p$ based on a sufficiently accurate approximation of $\|K \times 1\|_1$.

This relatively simple reduction gives strong running time lower bounds in the moderate/high dimensional regime. In the original paper, \cite{BIS17} shows that when $m = \Omega(\log n)$ and $B = \Omega(\log n)$, it requires $n^{2-o(1)}$ time to approximate the Gaussian KDE to $\eps = 2^{-\poly\log n}$. This result is later improved by Alman and Aggarwal \cite{AA22} (also in \cite{CS19} for KDE with relative error), who combine the same reduction with the hardness result of approximate BCP \cite{Rub18} and show that even approximating to accuracy $\eps = 1/\poly(n)$ requires $n^{2-o(1)}$ time.

However, we note that this reduction relies on a strong premise that $f(p)/f(p+1) > n^2$, i.e., the kernel function has \emph{rapid decay}. (The variant in \cite{AA22} relies on a similar condition. See Section \ref{sec:BISreduction} for more details.) This fails to hold for many kernels of interest, including all smooth kernels (such as the Rational Quadratic kernels and $t$-Student kernels) and small-scale Gaussian kernels with small $B < o(\log n)$.

To get around this barrier, Alman, Chu, Schild and Song \cite{ACSS20} extends the reduction of \cite{BIS17}, by solving \emph{multiple} KDE instances and, roughly speaking, combining their answers to extract \emph{more} information about the pairwise distances. Concretely, we define the \emph{distance vector} (in terms of two sets of points) and \emph{counting matrix} (in terms of the function $f$) as follows.
\begin{definition}
Let $X = \{x^{(1)}, \cdots, x^{(n)}\}, Y = \{y^{(1)}, \cdots, y^{(n)}\} \in \{0, 1\}^m$ be two sets of points. We define the distance vector $w = [\#\{(i, j): \|x^{(i)} - y^{(j)}\|_2^2 = p\}]_{p \in [m]}$, and for $\alpha_1, \cdots, \alpha_m \in [0, 1]$ define the counting matrix $M = [f(\alpha_{\ell} \cdot p)]_{\ell, p \in [m]}$.
\end{definition}
Consider the matrix-vector multiplication $M \times w$. By the same argument as in the reduction of \cite{BIS17}, we observe each entry in $\tilde w := M \times w$ can be computed by some KDE instance 
$K_{\ell} \times \bo$ ($K_\ell$ is associated with appropriately scaled $X$ and $Y$):
\[\tilde w[\ell] 
=\sum_{i=1}^n\sum_{j=1}^n f(\|\sqrt{\alpha_\ell}x^{(i)} - \sqrt{\alpha_\ell}y^{(j)}\|_2^2)
= \sum_{p=0}^m f(\alpha_\ell p) \cdot \#\left\{(i, j): \|x^{(i)} - y^{(j)}\|_2^2 = p\right\}.\]
Once $\tilde w$ is obtained from KDE subroutines, one can then approximate the distance vector by simply computing $M^{-1} \times \tilde w$. If this approximation to $w$ has (additive) error bounded by $1/3$, then a subsequent rounding step yields the exact distance vector, and automatically a solution to the BCP problem.

To analyze this reduction, we define a key quantity 
\[\tau(M) := \max_{0 \ne v \in \R^m} \frac{\|M^{-1}v\|_{\infty}}{\|v\|_{\infty}}.\] 
Suppose $\|K_\ell \times \bo\|_{\infty} \le \eps \|\bo\|_1$ for all $\ell \in [m]$. Then
\[\|M^{-1} \tilde w - w\|_\infty = 
\|M^{-1}(\tilde w - M w)\|_\infty \le \tau(M) \cdot \max_{\ell \in [m]}\|K_\ell \times \bo\|_1 \le \tau(M) \cdot n^2\eps.\]
Therefore, any algorithm that approximates KDE instances to accuracy $\eps = 1/(3n^2\tau(M))$ in $n^{2-\Omega(1)}$ time amounts to a truly subquadratic BCP algorithm and refutes SETH. (In this paper below, we slightly modify this approach to improve the dependence on $n$ in $\eps$ from quadratic to linear; see Section \ref{sec:techniques} for more details.)

To complete the hardness result, it remains to bound the quantity $\tau(M)$ as a function of $m$ and $B$. \cite{ACSS20} makes a generic statement relating $\tau(M)$ to the approximability of $f$ by low-degree polynomials. In particular, it is shown that for all three kernels -- Rational Quadratic kernel, $t$-Student kernel and small-scale Gaussian kernel -- performing $\Omega(\log n)$-dimensional KDE requires $n^{2-o(1)}$ time to achieve accuracy $\eps = 2^{-\poly\log n}$.

The proof of \cite{ACSS20} for this bound on $\tau(M)$ is highly technical, and requires new tools from analysis and linear algebra. For the sake of coherence, we defer an overview of the details to Section \ref{sec:techniques}. A main weakness of \cite{ACSS20} is that it only gives hardness for very small $\eps$, which is an inherent consequence of their approach to bounding $\tau(M)$. One of the main technical contributions of our paper, which we discuss in more detail shortly, is an improved approach to bounding $\tau(M)$ which yields exponentially better bounds on $\eps$ for many kernel functions of interest.

\subsection{Our contribution}
\label{sec:contribution}
In this work we give stronger negative results for the KDE problem, and pin down its complexity in each parameter regime. We mainly focus on the Gaussian kernel, and on two of the most used smooth kernels -- the Rational Quadratic kernel and the $t$-Student kernel. That said, our approach is general and would apply to any other kernel of interest after some calculations (See Section \ref{sec:discussion} for further discussion). To give a unified presentation, we formulate both the positive and negative results as upper bounds and lower bounds on $1/\eps$. More specifically, we will answer the following questions.
\begin{question}
Fix a kernel function $f$ and a parameter regime determined by $m = o(\log n)$ (resp. $\Theta(\log n)$, $\omega(\log n)$) and $B = o(\log n)$ (resp. $\Omega(\log n)$). What is the range of $1/\eps$ achievable in $n^{1+o(1)}$ time? What is the range of $1/\eps$ that requires $n^{2-o(1)}$ time?
\end{question}
For simplicity, in the following discussion we understand ``Easy'' as being achievable in $n^{1+o(1)}$ time, and understand ``Hard'' as requiring $n^{2-o(1)}$ time.

\paragraph{Gaussian kernel}
In the regime $m = \Theta(\log n), B = o(\log n)$, the polynomial method \cite{AA22} gives the best known positive result: Gaussian KDE is Easy when $1/\eps < (m/B)^{o(m)}$ (See also Section \ref{sec:polymethod}). The best known negative result due to \cite{ACSS20} establishes the Hardness of KDE when $1/\eps > 2^{\poly \log n}$. 
In this work we improve the negative result and show that the polynomial method is optimal up to a low-order $2^{O(m)}$ factor in $1/\eps$.
\begin{theorem} \label{gaussian1}
Assuming SETH, for every $q \in (0, 1)$, there exist $C_1, C_2 > 0$ such that when $m > C_1\log n$ and $1/\eps > (m/B)^{m} \cdot C_2^m $, $\mathsf{GaussianKDE}(n, m, B, \eps)$ cannot be solved in $O(n^{2-q})$ time.
\end{theorem}

In the low dimensional regime\footnote{We use $\log^*$ to denote the very slowly-growing iterated logarithm function.}, $c^{\log^*n} < m < o(\log n)/(\log \log n)$, the Fast Multipole Method has stood unchallenged for over three decades. Using this method, Gaussian KDE is Easy when $\log\log(1/\eps) < o(\log n)/m$. It is natural to conjecture that a substantial improvement is impossible. However, to the best of our knowledge, no previous hardness result was known for Gaussian kernel KDE in this regime.\footnote{For non-Lipschitz kernels, some hardness results for very low error in low dimensions were established in \cite{ACSS20}.} In this work we give the first negative result against the Fast Multipole Method, and in particular show that the $\log \log (1/\eps)$ achieved by the Fast Multipole Method is optimal up to a roughly logarithmic factor in $m$.
\begin{theorem} \label{gaussian2}
Assuming SETH, for every $q > 0$, there exist $C_1, C_1', C_2 > 0$ such that when $C_1^{\log^* n} < m < C_1'(\log n)/(\log\log n)$ and $\log \log (1/\eps) > (\log n)/m \cdot (\log m) \cdot C_2^{\log^* n}$, $\mathsf{GaussianKDE}(n, m, B, \eps)$ cannot be solved in $O(n^{2-q})$ time.
\end{theorem}

Apart from the two major improvements above, our techniques also lead to new results in other regimes. As a straightforward corollary of Theorem \ref{gaussian1}, we show $1/\eps > ((\log n)/B)^{\Omega(\log n)}$ is Hard for high-dimensional $m = \omega(\log n)$, small-scale $B = o(\log n)$ regime, improving the $1/\eps > 2^{\poly\log n}$ bound in \cite{ACSS20}. For the large scale regime, we note the hardness result in \cite{AA22} requires $(B/\log n)$ to tend to infinity alongside $(m/\log n)$. This inherent dependence between $B$ and $m$ is an inevitable consequence of the rapid decay condition. In comparison, as our new reduction is free of such restrictions, new hardness results can be developed as well in the regime where $B = \Theta(\log n)$ is fixed and only $m/\log n$ tends to infinity.

We summarize all the known upper and lower bounds for Gaussian KDE in Table~\ref{waffle}.

\begin{table}[htbp] 
\renewcommand{\arraystretch}{1.35}
\centering
\begin{tabular}{|c|c|c|}
\hline
& \begin{tabular}{c}small scale \\ $B = o(\log n)$\end{tabular}
& \begin{tabular}{c}large scale \\ $B = \Omega(\log n)$\end{tabular} \\ \hline

\begin{tabular}{c}low dimension \\ $c^{\log^* n} < m < o\left(\frac{\log n}{\log \log n}\right)$\end{tabular} 
& \multicolumn{2}{|c|}{
\begin{tabular}{c} 
Easy: $\log\log (1/\eps) < o((\log n)/m)$\\
Hard (new): $\log\log (1/\eps) > \tilde \Omega(\log n)/m)$ 
\end{tabular} } \\ \hline

\begin{tabular}{c}moderate dimension \\ $m = C\log n$\end{tabular} & 
\begin{tabular}{c} Easy: $1/\eps < (m/B)^{o(m)}$ \\
Hard (new): $1/\eps > \Omega(m/B)^m$
\end{tabular}
& \begin{tabular}{c} 
Easy: $1/\eps < n^{1 - q}$\\
Hard: $1/\eps > n^{C}$ for some $C > 1$
\end{tabular} \\ \hline

\begin{tabular}{c}high dimension \\ 
$m > \omega(\log n)$\end{tabular} 
& \begin{tabular}{c} 
Easy: $1/\eps < n^{1 - q}$\\
Hard (new): $1/\eps > ((\log n)/B)^{\Omega(\log n)}$
\end{tabular}
& \begin{tabular}{c} 
Easy: $1/\eps < n^{1 - q}$\\
Hard: $1/\eps > n^{C}$ for some $C > 1$
\end{tabular}
\\ \hline

\end{tabular}
\caption{Summary of known results for Gaussian KDE, incorporating our new Theorems~\ref{gaussian1} and \ref{gaussian2}. The new hardness in high dimensions follows from our Theorem~\ref{gaussian1} and a straightforward reduction from moderate to high dimension. The stated hardness results for large scale and moderate or high dimension were previously known~\cite{BIS17,AA22}, although we improve the constant $C$ in these cases.}\label{waffle}
\end{table}

\paragraph{Rational Quadratic kernel and $t$-Student kernel}
For the Rational Quadratic kernel $f(x)=1/(1+x)^\sigma$ and $t$-Student kernel $f(x) = 1/(1+x^\rho)$ parameterized by absolute constants $\sigma, \rho \ge 1$, we give similar lower bound results.
In the moderate to high dimensional regime $d = \Omega(\log n)$, the best known negative result is again that $1/\eps = 2^{\poly\log n}$ is Hard by \cite{ACSS20}. We show that this can be improved to $1/\eps = \poly(n)$. This parallels the improvement by \cite{AA22} for the large-scale Gaussian kernel from $1/\eps = 2^{\poly\log n}$ to $1/\eps = \poly(n)$ (and we recall that these kernels do not decrease quickly enough to prove this using the approach of \cite{BIS17,AA22}).

\begin{theorem} \label{thm:RQ1}
For Rational Quadratic kernel $f(x) = 1/(1+x)^{\sigma}$ and $t$-Student kernel $f(x) = 1/(1+x^\rho)$ parameterized by absolute constants $\sigma, \rho \ge 1$, the following holds.
Assuming SETH, for every $q > 0$, there exists $C_1, C_2 > 0$ such that if $m > C_1\log n, 1/\eps > n^{C_2}$, then $\KDE_f(n, m, B=1, \eps)$ cannot be solved in $O(n^{2-q})$. Here $C_2$ is dependent on $\sigma$ or $\rho$.
\end{theorem}
This result implies that KDE for Rational Quadratic kernel and $t$-Student kernel are strictly harder than Gaussian KDE: for $B = O(1)$ and $m = \Theta(\log n)$, Gaussian KDE is Easy when $1/\eps < m^{o(m)}$ whereas KDE for Rational Quadratic kernel and $t$-Student kernel are Hard for $1/\eps > 2^{\Omega(m)}$. Interestingly, this is in sharp contrast to the phenomenon in the study of KDE with relative error, where KDE for smooth kernels are seemingly easier to solve. We discuss this difference in detail in Section \ref{sec:discussion}.

In the low dimensional regime, we also prove negative results against the Fast Multipole Method. For both Rational Quadratic kernel and $t$-Student kernel, the Fast Multipole Method has similar bound $O(n\log^{O(m)}(n/\eps))$ on running time. We complement this algorithm with a matching lower bound up to a $\tilde O(\log m)$ factor in $\log \log(1/\eps)$.
\begin{theorem}
For Rational Quadratic kernel $f(x) = 1/(1+x)^{\sigma}$ and $t$-Student kernel $f(x) = 1/(1+x^\rho)$ parameterized by absolute constants $\sigma, \rho \ge 1$, the following holds.
Assuming SETH, for every $q > 0$, there exist $C_1, C_1', C_2 > 0$ such that when $C_1^{\log^*n} < m < C_1'(\log n)/(\log\log n)$ and $\log\log(1/\eps) > (\log n)/m \cdot (\log m)\cdot C_2^{\log^*n}$, $\KDE_f(n, m, B=1, \eps)$ cannot be solved in $O(n^{2-q})$ time. Here $C_2$ is dependent on $\sigma$ or $\rho$.
\end{theorem}

\subsection{Techniques}
\label{sec:techniques}
As discussed in Section \ref{sec:prevlb}, \cite{ACSS20} established an upper bound on $1/\eps$ which hinges on a key quantity $\tau(M)$, and the central ingredient of their reduction is a bound on this quantity.
Therefore we start by sketching the proof \cite{ACSS20} developed for this bound. By definition,
\[\tau(M) = \max_{0 \ne v \in \R^m} \frac{\|M^{-1} v\|_{\infty}}{\|v\|_{\infty}}
=\max_{\|v\|_\infty = 1} \|M^{-1} v\|_\infty
\le \max_{\|v\|_\infty = 1} \max_{t \in [m]} \sum_{s=1}^{m} |M^{-1}[t,s]||v[s]| 
\le m\max_{t,s \in [m]}|M^{-1}[t, s]|.\]
By Cramer's rule, we write
\[M^{-1}[t, s] = \frac{\det(M_{t-}^{s-})}{\det (M)}\]
where $M_{t-}^{s-}$ is the matrix obtained by removing the $s$-th row and $t$-th column of $M$.
Thus it suffices to bound $\det(M)$ and $\det(M_{t-}^{s-})$.
We here make use of a common matrix decomposition technique in the study of the polynomial method.
If $f$ has Taylor series $f(x) = \sum_{k=0}^{\infty} \frac{f^{(k)}(0)}{k!}x^k$ convergent over $[0, 1]$, then
\[\det(M) = \det\bigg[f(\alpha_s\beta_t)\bigg]_{s, t \in [m]}
= \det\bigg[\sum_{k=0}^{\infty}\frac{f^{(k)}(0)}{k!}\alpha_s^k\beta_t^k\bigg]_{s, t\in [m]}
= \det\left(\bigg[\alpha_s^k \cdot \frac{f^{(k)}(0)}{k!}\bigg]_{m \times \N} 
\times \bigg[\beta_t^k\bigg]_{\N \times m}\right).\]
One tool for computing determinants of the form $\det(A \times B)$, where $A$ and $B$ are rectangular matrices, is the Cauchy-Binet formula (See Section \ref{subsec:CauchyBinetTool} for details), which gives
\begin{equation}
\label{eq:CBexpansion}
\det(M) = \sum_{0 \le n_1 < \cdots < n_m} \left(\prod_{k=1}^m \frac{f^{(n_k)}(0)}{n_k!}\right)
\det\bigg[\alpha_s^{n_k}\bigg]_{s, k\in [m]}\det\bigg[\beta_t^{n_k}\bigg]_{t, k \in [m]}. 
\end{equation}
Observing that the determinants involved are effectively $(m!)$-term polynomials in $\alpha$ and $\beta$, \cite{ACSS20} then views the entire sum as a power series and applies a standard (yet technical) analysis to derive a bound on $\tau(M)$.

In this work, we extend this approach in four aspects.
\paragraph{Direction 1: Schur polynomials} 
First, we improve on the analysis of the series (\ref{eq:CBexpansion}). Although this series is a rather concrete representation of $\det(M)$, the analysis of the $(m!)$-term polynomials and infinite sum is still a strenuous task and tends not to lead to tight bounds. In this work, we make further inspections of the structure of series (\ref{eq:CBexpansion}) and observe that all the determinants $\det\left[\alpha_s^{n_k}\right]_{s, k\in [m]}, \det\left[\beta_t^{n_k}\right]_{t, k \in [m]}$ have a special structure -- they are known as \emph{generalized Vandermonde matrices}. Such matrices have been extensively studied in Algebraic Combinatorics under the name \emph{Schur polynomials}, and are central to the theory of symmetric polynomials. For the classical theory and application of Schur polynomials we point to Chapter 7 of \cite{Stanley}. In recent years, Schur polynomials have also found various applications in computer science, such as in quantum computation \cite{HRTS00, OW15} and geometric complexity theory \cite{geocomp}.

One key property of Schur polynomials lies in its two equivalent definitions. The algebraic definition by Cauchy establishes connection between Schur polynomials and generalized Vandermonde matrices, while the combinatorial definition by Littlewood gives a concrete specification of the coefficients of Schur polynomials.
\begin{definition}[Schur polynomials]
Let $m>0$ be an integer and $\lambda_1 \le \cdots \le \lambda_m$ be positive integers.
We define the Schur polynomial $s_\lambda$ on variables $(u_1, \cdots, u_m)$ by
\[s_{\lambda}(u) = s_{\lambda}(u_1, \cdots, u_m)
=\frac{\det[u^{\lambda_k + (k-1)}_{i}]_{i, k \in [m]}}
{\prod_{j > i}(u_j - u_i)}. \tag{Cauchy}\]
Equivalently, the Schur polynomial $s_{\lambda}(u)$ can be defined by
\[s_{\lambda}(u) = \sum_{T \in \mathcal T} \prod_{i=1}^m u_i^{\mathrm t(T)_i}, \tag{Littlewood}\]
where $\mathcal T$ is the set of all semi-standard Young tableaux of shape $\lambda$ on alphabet $[m]$, and $\mathrm t(T) \in \N^m$ is the type of $T$. (See Section \ref{sec:Schur} for the definition of Young tableaux and associated parameters.)
\end{definition}

Based on this property, we are able to represent $\det\left[\alpha_s^{n_k}\right]$ and $\det\left[\beta_t^{n_k}\right]$ by ``neater'' polynomials whose nonzero coefficients are uniformly 1, and whose monomials can be enumerated using some combinatorial objects. One can thereby obtain stronger bounds through more straightforward analysis. Moreover, many known results regarding the asymptotic behavior of Schur polynomials also turn out useful in providing guidance on proof strategies.

\paragraph{Direction 2: Special counting matrices}
We also observe that for many kernels of interest, the counting matrix $M$ itself has a special structure. For the $t$-Student kernel $f = 1/(1+x^\rho)$, its associated counting matrix 
$M_f = [1/(1+\alpha_s^\rho\beta_t^\rho)]_{s, t\in[m]}$ is known as a (scaled) Cauchy matrix, as is $M^{s-}_{t-}$.
For the Gaussian kernel $f = e^{-x}$, the associated counting matrix $M_f = [e^{\alpha_s\beta_t}]_{s,t\in [m]}$ is a Vandermonde matrix, and $M^{s-}_{t-}$ is (a relatively simple example of) a generalized Vandermonde matrix.
Closed-form formulas are known for determinants of such special matrices. One may thus get around the Cauchy-Binet expansion and bounds can be deduced via a direct argument.

\paragraph{Direction 3: Grouping vector pairs}
Regarding the reduction per se, we show the reduction in \cite{ACSS20} can be modified so that $\eps = 1/(3n \tau(M))$ suffices, in contrast to the aforementioned $\eps = 1/(3n^2\tau(M))$ lower bound. 
The main idea is to perform the reduction on $\{x^{(i)}\} \times Y$ for each $x^{(i)} \in X$ separately. Note that for fixed $i \in [n]$,
\[(K\times \bo)[i] = \sum_{j \in [n]} f(\|x^{(i)} - y^{(j)}\|_2^2)
=\sum_{p=0}^m f(p) \cdot \#\left\{j: \|x^{(i)} - y^{(j)}\|_2^2 = p\right\}.\]
By the same argument as before, one now recovers the components $w^i = [\#\{j \in [m]: \|x^{(i)} - y^{(j)}\|_2^2 = p\}]_{p \in [m]}$ of $w$ for respective $i\in [m]$. We note, in this new reduction, it suffices to approximate single entries $(K \times \bo)[i]$, which is arguably a simpler task than approximating $\|K \times \bo\|_1$, as an error $n$ times as large may accumulate in the latter case. Formalizing this idea in Section \ref{sec:mainreduction}, we successfully shave a factor of $n$.

This improvement on the dependence of $n$ in addition raises an interesting question. In the high dimensional regime, \cite{PT20} showed that for any positive definite kernel $f$, one can compute $\KDE_f(n, m, \eps, B)$ in truly subquadratic time with accuracy $1/\eps < n^{1-\delta}$ for any constant $\delta > 0$.
Against this algorithm, our improved reduction produces a lower bound for $1/\eps > \Omega(n) \cdot \tau(M)$, leaving a gap of simply $\tau(M)$.
This brings within reach a potential tightness result: the optimality of \cite{PT20} can be (in part) established as long as one can bound $\tau(M) = O(1)$ for some positive definite kernel $f$.
Such a result is arguably hard to obtain from the previous lower bounds by either \cite{BIS17} or \cite{ACSS20}.

\paragraph{Direction 4: Low-dimensional BCP}
To extend the negative result to the low-dimensional regime, we combine our main reduction with a variant of the BCP problem. Williams \cite{Wil18} and Chen \cite{ChenMinIP} showed that the BCP problem for vectors with \emph{integer entries} remains hard even in extremely low dimensions $d = 2^{O(\log^* n)}$. (See Section \ref{sec:hardproblems} for a formal statement). With slight modification, our main reduction can use KDE subroutines to recover the distance vector for not only datapoints in $\{0, 1\}^m$ but those in $\Z^m$ (though a larger counting matrix is required). A hardness result for KDE in low dimensions thus follows from similar analysis. Moreover, looking into the proofs of \cite{Wil18} and \cite{ChenMinIP}, we notice that they effectively showed a stronger trade-off between the dimension of vectors and magnitude of vector entries.
Translated into the setting of KDE, this is a trade-off between dimension $m$ and approximation error $\eps$.
In Section \ref{sec:WilliamsChen} we formalize the trade-off between dimension and magnitude and give detailed calculations.

\subsection{Discussion}
\label{sec:discussion}
\paragraph{Additive vs. Relative Error} In the previous discussion we focused mainly on the KDE problem with additive error. In recent years, much effort has also been dedicated to algorithm design in the setting with relative error, primarily in the moderate to high dimensional regime $d = \Omega(\log n)$. In this setting, the running time of KDE algorithms normally depends not only on the relative error parameter $\eps_R$ but also on a lower bound of the kernel value $\mu = \min_{x \in [0, B]}f(x)$. 
The folklore random sampling algorithm runs in time $O(n\eps_R^{-2}\mu^{-1})$. For the Gaussian kernel, Charikar and Siminelakis \cite{CS17} made the first major improvement by designing a $O(n\eps_R^{-2}\mu^{-0.5})$-time algorithm using a LSH-based Importance Sampling scheme. Later Charikar et al. \cite{CKNS20} presented an improved implementation of Importance sampling that achieves $O(n\eps_R^{-2}\mu^{-0.173})$ running time. 
For smooth kernels (including the Rational Quadratic kernel and $t$-Student kernel), the first non-trivial improvement was due to Backurs et al. \cite{BCIS18}, who presented an algorithm running in $n\eps_R^{-2}\poly\log(\mu^{-1})$ time using tree-based space partitioning techniques. Recently, Charikar, Kapralov and Waingarten \cite{CKW24} combined this result with the discrepancy based sampling scheme by Phillips and Tai \cite{PT20} and achieved $n \eps_R^{-1}\poly\log(\mu^{-1})$ running time, improving the dependence on $\eps_R$.

Interestingly, our new lower bounds in Section \ref{sec:contribution} exhibits a sharp contrast between the additive and relative error setting. The best known KDE algorithms stated above suggest that KDE for smooth kernels are likely easier than that for Gaussian-like kernels in the relative error setting. However, comparing between Theorem \ref{gaussian1} and Theorem \ref{thm:RQ1}, we observe an opposite trend. For example, for $B = O(1)$ and $m = \Theta(\log n)$, Gaussian KDE is Easy when $1/\eps < m^{o(m)}$ whereas KDE for Rational Quadratic kernel and $t$-Student kernel are Hard even for $1/\eps = 2^{\Omega(m)}$. This difference suggests that the discrepancy between two formulations is likely inherent and they should be treated with respective care.

\paragraph{Dynamic vs. Batched KDE} As we see from the reduction of \cite{BIS17} and \cite{ACSS20}, the BCP problem naturally reduces to the batched version of KDE, and thereby we take batched KDE as the primary formulation in this work for simplicity. In the literature, the KDE problem is equally often phrased in its dynamic version, e.g., in \cite{CS17, PT20, CKW24}. In the dynamic KDE problem for kernel $k(x, y)$, one is given a dataset $X \subset \R^m$ and a vector $w \in \R^n$, and asked to design a data structure $\mathcal A$ that preprocesses $X$ and outputs an approximation to the Kernel Density $\sum_{x\in X}k(x, q)w[x]$ for each query point $q \in \R^m$. Given a KDE data structure, one can easily build a batched KDE algorithm in time $T$(preprocessing) + $n \cdot T$(query). Hence any hardness result proved for batched KDE automatically holds for dynamic KDE as well.
It is not clear whether there is a reduction in the reverse direction, and it remains an open problem to determine whether the batched version is strictly easier. Nonetheless, all the known algorithms including the Fast Multipole method, Polynomial method and sampling-based methods, are data structures or can be modified to solve the dynamic problem.

\paragraph{Other Kernels} We note that our reduction also provides hardness results for more kernels than just the three we study in detail. In the previous discussion, we showed that our main reduction from Hamming BCP yields negative results for KDE for radial kernels, i.e., kernels of the form $k(x, y) = f(\|x - y\|_2^2)$ for some $f$. As $\|x-y\|_2^2 = \|x-y\|_1$ for $x, y \in \{0, 1\}^n$, this same reduction can also be used to demonstrate hardness for kernels of the form $f(\|x-y\|_1)$ such as the Laplacian kernel $k(x, y) = \exp(-\|x-y\|_1)$ studied in \cite{BIW19}. If we instead use our approach to reduce from the Hamming or $\Z$ Orthogonal Vectors problem, on the other hand, then we get hardness results for kernels of the form $f(\ip{x}{y})$ such as the Partition function $k(x, y) = \exp(\ip{x}{y})$ studied in \cite{CS19}.

On another note, we in Section \ref{sec:pdkde} show that all the positive definite kernels can be related to an absolutely monotonic function, whose associated $\tau(M)$ is a ratio of two sums of Schur polynomials with \emph{positive} coefficients. Standard analysis of power series and basic analytic properties of Schur polynomials are usually sufficient to give a reasonably good bound on sums of this simple form, from which complexity results immediately follow. In contrast to \cite{BIS17} which requires the kernel to be rapidly decreasing, and to \cite{ACSS20} which requires certain knowledge about the approximate degree of $f$, our method is more suitable for use when the Taylor coefficients of $f$ are easy to describe.

\paragraph{Expansion in terms of Schur polynomials}
Implicitly, the idea of representing the determinant of counting matrices by Schur polynomials has been explored in other contexts which helped inspire our proof here. One incarnation appeared in the characterization of positivity preservers by Khare and Tao \cite{KT21}. We say a function $f:\R_{\ge 0} \rightarrow \R$ is a positivity preserver if, for any matrix $A = [a_{jk}]_{j, k \in [N]}$ which is positive semidefinite, the matrix $f[A] := [f(a_{jk})]_{j, k \in [N]}$ is also positive semidefinite. If we consider a rank-1 positive semidefinite matrix $A = uu^T$, then the entries in $f[A]$ are $f(u_iu_j)$. This is a special case of what we call a counting matrix associated with $(f, \alpha = \beta = u)$. By expanding $\det(f[A])$ into a sum of Schur polynomials in effectively the same way as we do, they showed that, informally speaking, the Taylor coefficients must be all nonnegative to ensure the positivity of $\det(f[A])$.

Another example occurs in the study of the Schur measure, a probability measure on partitions which is used in the study of many random combinatorial objects.\footnote{We recommend returning to this paragraph after reading Section \ref{sec:cauchybinet}.} Consider, for instance, the Cauchy kernel (which corresponds to $C(u) = 1/(1-u)$). We can calculate that the representation of its counting matrix by Schur polynomials is $\det(M_C) = \sum_{\lambda \in \N^{m}_{\le}}s_{\lambda}(\alpha)s_{\lambda}(\beta)$. This quantity is known to equal $\prod_{i,j} (1 - \alpha_i \beta_j)^{-1}$ via the Cauchy identity for Schur polynomials, and it defines the Schur measure on all length-$m$ partitions $\lambda \in \N^{m}_{\lambda}$ by $\mathbb P_{\alpha, \beta}(\lambda) = \frac{s_{\lambda}(\alpha)s_{\lambda}(\beta)}{\det(M_C)}$.
A common task concerning the Schur measure is to find a closed-form expression for the expectation 
$\Ex_{\mathbb P}[\phi(\lambda)]$ of a given function $\phi$.
One major result along this line is due to Okounkov \cite{Okounkov01} (see also the lecture notes \cite{SchurMeasure}),
who studied the function $\phi_A = \ind[A \subseteq \{\lambda_i + \delta_i: i\in [m]\}]$ for any fixed set $A$ of size $r$, and showed there exists a kernel $K: \R^k \times \R^k \rightarrow \R$ and a vector $u\in \R^r$ such that $\Ex_{\mathbb P}[\phi_A(\lambda)] = \det[K(u_i, u_j)]_{i, j \in [r]}$. In other words, the expectation $\Ex_{\mathbb P}[\phi(\lambda)]$ can be computed by the determinant of a ``counting matrix'' associated with $(K, \alpha = \beta = u)$.
From this perspective, our expansion in terms of Schur polynomials effectively produces a similar result,
by equating the determinant of a counting matrix $\det(M_f)$ with an expectation $\Ex_{\mathbb P}[\phi_T(\lambda)]$ where $\phi_T(\lambda) = \prod_{i=1}^m \frac{f^{(\lambda_i + \delta_i)}(c)}{(\lambda_i + \delta_i)!}$ is a product of Taylor coefficients. In light of this connection, we hope future work can use more kernels studied in this context to bound the complexity of associated KDE problems.

\section{Preliminaries}
\subsection{Notation and assumptions}
We let $N$ denote the set of all natural numbers $\{0, 1, 2, \cdots\}$, and $\Z_{+}$ denote all positive integers $\{1, 2, \cdots\}$. We let $[n]$ denote the $\{1, \cdots, n\}$ for $n \in \Z_{+}$.
For vector $v \in \R^m$ and $s \in [m]$, we let $v^{s-} \in \R^{m-1}$ denote the vector obtained by removing the $s$-th entry in $v$. Similarly, for a matrix $A \in \R^{m \times m}$ and $s, t \in [m]$, we let $A^{s-}_{t-} \in \R^{(m-1) \times (m-1)}$ denote the matrix obtained by removing the $s$-th row and $t$-th column in $A$. 

The iterated logarithm function $\log^*(n)$ is defined recursively as follows.
\[\log^*(n) = \left\{\begin{array}{ll}0 & n \le 1; \\ \log^*(\log n)+1 & n > 1.\end{array}\right.\]

We assume throughout this work that for a given kernel $k: \R^k \times \R^k \rightarrow \R$ and inputs $x, y \in \R^m$, the value $k(x, y)$ is (exactly) computable in constant time. This assumption is commonly adopted in the literature, e.g.,~\cite{CS17, CKW24}.

\subsection{Reduction from large domain $[0, B]$ to normal domain $[0, 1]$}
For the simplicity of analysis, we will integrate parameter $B$ into the kernel function $f$ via the following lemma.
\begin{lemma}
For integers $n, m \in \Z_{\ge 1}$, real numbers $B, \eps \in \R_{>0}$,
and function $f: [0, B] \rightarrow \R$,
$\KDE_f(n, m, B, \eps)$ can be solved in $T(n, m, \eps, B)$ time if and only if
$\KDE_{f_B}(n, m, \eps, 1)$ can be solved in $T(n, m, \eps, B)$ time,
where $f_B: [0, 1] \rightarrow \R$ is defined by $f_B(x) = f(Bx)$.
\end{lemma}
\begin{proof}
We first prove the ``if'' direction.
Given vectors $x^{(1)}, \cdots, y^{(n)} \in \R^m$, 
suppose $\|x^{(i)} - y^{(j)}\|_2^2 \le B$ for all $i, j \in [n]$. Then
\[f(\|x^{(i)} - y^{(j)}\|_2^2) = f_B(\|B^{-1/2} x^{(i)} - B^{-1/2} y^{(j)}\|_2^2).\]
Now the vectors $\tilde x^{(i)} := B^{-1/2} x^{(i)}, \tilde y^{(j)} := B^{-1/2} y^{(j)}, i, j \in [n]$ satisfy $\|\tilde x^{(i)} - \tilde y^{(j)}\|_2^2 \le 1$. Solving $\KDE_{f_B}(n, m,$ $\eps, 1)$ on $\tilde x^{(1)}, \cdots, \tilde y^{(n)}$ yields precisely the result for 
$\KDE_{f}(n, m, \eps, B)$ on $x^{(1)}, \cdots, y^{(n)}$.
Proof for the converse is analogous.
\end{proof}

In light of this reduction, we in the following discussion will focus on KDE problems $\KDE_f(n, m, \eps) := \KDE_f(n, m, \eps, B=1)$ on the normal domain $[0, 1]$, where $f$ can be dependent on or independent of $B$.

\subsection{SETH and known hard problems}
\label{sec:hardproblems}

The lower bounds in this paper will be predicated on the popular Strong Exponential Time Hypothesis (SETH) from fine-grained complexity theory.

\begin{conjecture}[SETH~\cite{impagliazzo2001complexity}]
For every $\delta>0$ there is a positive integer $k$ such that $k$-SAT instances with $n$ variables cannot be solved in time $O(2^{n(1-\delta)})$ by a randomized algorithm.
\end{conjecture}

We will also use the closely-related Orthogonal Vectors problem.

\begin{problem}[Orthogonal Vectors]
\label{prob:OV}
$\emph{\textsf{OV}}(n, m)$: Given two sets $A = \{x^{(1)}, \cdots, x^{(n)}\}, B = \{y^{(1)}, \cdots, y^{(n)}\} \subset \{0, 1\}^m$, determine whether there are $i,j \in [n]$ with $\langle x^{(i)} - y^{(j)} \rangle$ = 0.
\end{problem}

\begin{conjecture}[Orthogonal Vectors Conjecture]
For every $\delta > 0$, there is $c \ge 1$ such that \emph{\textsf{OV}}$(n, m)$ cannot be solved in $O(n^{2-\delta})$ time on instances with $m = c\log n$.
\end{conjecture}

It is known that SETH implies the Orthogonal Vectors Conjecture~\cite{williams2005new}.

We now introduce several variants of the Bichromatic Closest Pair problem. 
\begin{problem}[Hamming (Exact) Bichromatic Closest Pair]
\label{prob:HBCP}
$\emph{\textsf{Hamming-BCP}}(n, m)$: Given two sets $A = \{x^{(1)}, \cdots, x^{(n)}\}, B = \{y^{(1)}, \cdots, y^{(n)}\} \subset \{0, 1\}^m$, compute $\min_{i, j \in [n]} \|x^{(i)} - y^{(j)}\|_2^2$.
\end{problem}
\begin{theorem}[\cite{AW15}]
Assuming SETH, for every $q \in (0, 1)$, there exists $C > 0$ such that if $m > C\log n$, then $\mathsf{Hamming}$-$\mathsf{BCP}(n, m)$ cannot be solved in time $O(n^{2-q})$ for any constant $q>0$.
\end{theorem}

Similarly, one can define the Hamming approximate BCP problem and its decision version.
\begin{problem}[Hamming Approximate BCP]
\label{prob:HABCP}
$\emph{\textsf{Hamming-Apx-BCP}}(n, m, \mu)$:
Given two sets $A, B$ as in Problem \ref{prob:HBCP} as well as $\mu \in \R_+$, output $d \in \R$ such that $\min_{i, j \in [n]} \|x^{(i)} - y^{(j)}\|_2^2 \le d \le (1 + \mu) \min_{i, j \in [n]} \|x^{(i)} - y^{(j)}\|_2^2$.
\end{problem}
\begin{theorem}[\cite{Rub18}] \label{bcp1}
Assuming SETH, for every $q > 0$, there exist $C > 0, \mu > 0$ such that if $m > C\log n$, then $\emph{\textsf{Hamming-Apx-BCP}}(n, m, \mu)$ cannot be solved in time $O(n^{2-q})$ for any constant $q>0$.
\end{theorem}

In the low dimensional regime $m = o(\log n)$, we consider the hardness of the $\ell_2$ BCP problem.
\begin{problem}[$\ell_2$ (Exact) Bichromatic Closest Pair]
$\ell_2\emph{\textsf{-BCP}}(n, m, D)$: Given two sets $A = \{x^{(1)}, \cdots, x^{(n)}\}, B = \{y^{(1)}, \cdots, y^{(n)}\} \in \Z^m$ such that $\max_{i, j \in [n]} \|x^{(i)} - y^{(j)}\|_2^2 \le D$, compute $\min_{i, j \in [n]} \|x^{(i)} - y^{(j)}\|_2^2$.
\end{problem}
\begin{theorem}[\cite{ChenMinIP}] \label{bcp2}
Assuming SETH, for every $q > 0$, there exists $C_1, C_2 > 0$ such that if $m > C_1^{\log^* n}$ and $D > m^{C_2^{\log^* n} \cdot (\log n)/m}$, then $\ell_2$\emph{\textsf{-BCP}}$(n, m, D)$ cannot be solved in time $O(n^{2-q})$ for any constant $q>0$.
\end{theorem}

To unify the hardness results for BCP in different dimension regimes, we view the Hamming BCP problem 
as an $\ell_2$ BCP problem with $D = m$. 
Formally, we combine Theorem~\ref{bcp1} and Theorem~\ref{bcp2} as follows.
\begin{theorem} \label{thm:mainbcp}
Assuming SETH, for every $q > 0$, there exists $C > 0$ such that $\ell_2$\emph{\textsf{-BCP}}$(n, m, D)$
cannot be solved in time $O(n^{2-q})$ for any constant $q>0$ if either of the following holds: (1) $m > C \log n, D = m$, or (2) $m > C^{\log^* n}, D > m^{C^{\log^* n} \cdot (\log n)/m}$.
\end{theorem}

\subsection{Kernels of interest}
In this work we focus primarily on three kernels $k(x, y) = f(\|x - y\|_2^2)$:
\begin{itemize}
\item Gaussian kernel $f(x) = e^{-x}$;
\item Rational Quadratic kernel $f(x) = 1/(1+x)^\sigma$ for $\sigma \ge 1$ a parameter;
\item $t$-Student kernel $f(x) = 1/(1+x^{\rho})$ for $\rho \ge 1$ a parameter.\footnote{$t$ in the name of the kernel in principle should be the name of the parameter. We here use $\rho$ as parameter while keeping the name \emph{$t$-Student kernel} unchanged.}
\end{itemize}
Rational Quadratic kernel and $t$-Student kernel are two typical kernels with mild decay (as opposed to the rapid decay of Gaussian kernel). This property is abstracted by Backurs et al. \cite{BCIS18} in the definition of a smooth kernel. Here we only focus on decreasing radial kernels.
\begin{definition}[smooth kernel]
A decreasing radial kernel $k(x, y) = f(\|x - y\|_2^2)$ is $(L, t)$ smooth if for any $0 < a < b$,
\[\frac{f(a)}{f(b)} \le L \left(\frac ba \right)^t.\]
\end{definition}
By calculation one can verify that Rational Quadratic kernel and $t$-Student kernel are respectively $(1, 1)$- and $(1, \rho)$-smooth.

\paragraph{Positive definite kernels}
Most commonly-studied kernels are positive definite. We will find that our lower bound approach takes a particularly nice form for such kernels.

\begin{definition}[Positive definite kernel]
A kernel $k: \R^m \times \R^m \rightarrow \R$ is positive definite if for any $n$ points $x_1, \cdots, x_n \in \R^m$, the Gram matrix $G = [k(x_i, x_j)]_{i, j \in [n]}$ is always positive definite.
\end{definition}
For radial kernels, we have the following concise characterization of positive definite kernels.
\begin{definition}
Let $f:\R_{\ge 0} \rightarrow \R$ be a real function. We say $f$ is absolutely monotone if $G \in C^{\infty}(\R_{\ge 0})$ and $f^{(k)}(t)$ for all $k \in \N$ and $t \ge 0$, and we say $f$ is completely monotone if $G \in C^{\infty}(\R_{\ge 0})$ and $(-1)^{k} \cdot f^{(k)}(t) \ge 0$ for all $k \in \N$ and $t \ge 0$.
\end{definition}
\begin{theorem}[Schoenberg's characterization] \label{schoenberg} Let $f:\R_{\ge 0} \rightarrow \R$ be a real function. Then the kernel $k(x, y) = f(\|x-y\|_2^2)$ is positive definite if and only if $f$ is completely monotone on $\R_{\ge 0}$.
\end{theorem}

\subsection{Tools from linear algebra}
\label{subsec:CauchyBinetTool}

As in prior work, we will make use of the Cauchy-Binet formula.

\begin{lemma}[Cauchy-Binet formula]
Let $k>0$ be an integer, and for functions $A: [k] \times \N \rightarrow \R$ and $B: \N \times [k] \rightarrow \R$, define the matrix $C \in \R^{k \times k}$ by, for $i, j \in [k]$, $C[i, j] = \sum_{\ell=0}^{\infty} A[i, \ell]B[\ell, j]$.
If the sum converges absolutely for all $i, j$, then
\[\det(C) = \sum_{1 \le \ell_1 < \cdots < \ell_k}
\det(A[\ell_1, \cdots, \ell_k]) \cdot \det(B[\ell_1, \cdots, \ell_k]).\]
Here $A[\ell_1, \cdots, \ell_k]$(resp. $B[\ell_1, \cdots, \ell_k]$) is the $k \times k$ matrix obtained from $A$ (resp. $B$) by taking the columns (resp. rows) $\ell_1, \cdots, \ell_k$.
\end{lemma}

\section{Main reduction from BCP}
\label{sec:mainreduction}
Most of our new lower bounds are based on the reduction below from Hamming or $\ell_2$ Exact Bichromatic Closest Pair to KDE. This reduction generalizes a framework developed in \cite{ACSS20} to accommodate different parameter regimes. In this section we will give the outline of the reduction and establish an upper bound on $1/\eps$ determined by the quantity $\tau(M)$.

\begin{definition}
\label{def:countingmatrix}
Let $D > 0$ be an integer. Fix vectors $\alpha, \beta \in \R^D$ and $c \in \R$. Suppose $c + \alpha_{\ell}\beta_{p} \in [0, 1]$ for all $\ell, p \in [D]$. Then for function $f: [0, 1] \rightarrow \R$, the counting matrix $M = M(D;f, \alpha, \beta)$ is a $D \times D$ matrix defined by $M[\ell, p] = f(c + \alpha_{\ell}\beta_p), \ell, p \in [D]$.
\end{definition}
\begin{theorem}
Let $\alpha \in \R^D$ be a fixed vector and $\beta \in \R^D$ be the identity vector defined by $\beta_p = p$. If $\KDE_f(n,m,\eps)$ can be solved in $T(n, m, \eps)$ time, then $\ell_2$\emph{\textsf{-BCP}}$(n, m, D)$ with $m = n^{o(1)}, D = n^{o(1)}$ can be solved in $T(n, m+1, (3n\tau(M))^{-1}) \cdot n^{o(1)} + n^{1 + o(1)}$ time, where $M$ is the $D \times D$ counting matrix associated with $f, \alpha, \beta$ and
\[\tau(M) = \max_{0 \ne b \in \R^{D}} \frac{\|M^{-1}b\|_{\infty}}{\|b\|_{\infty}}.\]
\end{theorem}

\begin{proof}
Given two sets $X = \{x^{(1)}, \cdots, x^{n}\}, Y = \{y^{(1)}, \cdots, y^{(n)}\} \subseteq \Z^m$ such that $\max_{i, j \in [n]} \|x^{(i)} - y^{(j)}\|_2^2 \le D$, let $W \in \N^{D \times n}$ denote the \emph{distance count matrix} defined by $W[p, i] = \#[j \in [n]: \|x^{(i)} - y^{(j)}\|_2^2 = p]$. Then the matrix product $U = M \times W$ gives
\begin{align}
U[\ell, i] &= \sum_{p=1}^D M[\ell, p] \cdot W[p, i] \notag\\ 
&=\sum_{p=1}^D f\big(c + \alpha_\ell\beta_p\big) 
\sum_{j \in [n]} \ind\left[\|x^{(i)} - y^{(j)}\|_2^2 = p\right] \notag\\
&=\sum_{j \in [n]} f\left(c + \alpha_\ell \cdot \beta\left[\|x^{(i)} - y^{(j)}\|_2^2\right]\right). \label{eq:kernelsum} 
\end{align}
We note that when $\beta$ is the identity vector, the summation (\ref{eq:kernelsum}) can be formulated as a KDE instance.
More specifically, let $\tilde x_{\ell}^{(i)}, \tilde y_{\ell}^{(j)} \in \R^{m+1}$ be defined by $\tilde x_{\ell}^{(i)}[k] = \sqrt{\alpha(\ell)} x^{(i)}[k]$ if $k \in [m]$ and $\tilde x_{\ell}^{(i)}[m+1] = \sqrt c$;
$\tilde y_{\ell}^{(j)}[k] = \sqrt{\alpha(\ell)} y^{(j)}[k]$ if $k \in [m]$ and $\tilde y_{\ell}^{(j)}[m+1] = 0$. Then
\[U[\ell, i] = \sum_{j \in [n]} f(\|\tilde x_{\ell}^{(i)} - \tilde y_{\ell}^{(j)}\|_2^2).\]
Therefore we have the following algorithm for $\ell_2$\textsf{-BCP}$(n, m, D)$ using $\KDE_f$ as a subroutine. 
On input $X = \{x^{(1)}, \cdots, x^{(n)}\}, Y = \{y^{(1)}, \cdots, y^{(n)}\}$ such that $\max_{i, j \in [n]} \|x^{(i)} - y^{(j)}\|_2^2 \le D$:
\begin{enumerate}
\item For $\ell \in [D]$, construct vectors $\tilde x_{\ell}^{(i)}, \tilde y^{(j)}_{\ell}, i, j \in [n]$.
Then approximate the $\ell$-th row of $U$: $\hat U[\ell] \approx U[\ell] = K_{\ell} \times \bo$ using the $\KDE_f$ oracle.
\item Compute $\hat W = M^{-1} \times \hat U$, and round each entry to the closest integer.
\end{enumerate}

We claim that if we call the $\KDE_f$ subroutine with $\eps = (3n\tau(M))^{-1}$, then the distance count matrix $W$ is exactly recovered after the rounding step.
Indeed, for fixed $i \in [n]$, letting $W[\cdot, i], U[\cdot, i]$ respectively denote the $i$-th column of matrix $W$ and $U$, we have
\[\|\hat W[\cdot, i] - W[\cdot, i]\|_{\infty}
=\|M^{-1}(\hat U[\cdot, i] - U[\cdot, i])\|_{\infty}
\le \tau(M) \cdot \|\hat U[\cdot, i] - U[\cdot, i]\|_{\infty}.\]
If the $\KDE_f$ subroutine guarantees that $\|\hat U[\ell] - U[\ell]\|_{\infty} \le (3n\tau(M))^{-1} \cdot \|\bo\|_{1} = (3\tau(M))^{-1}$ for all $\ell \in \{1, \cdots, D\}$, then the entry-wise difference between $\hat W$ and $W$ is bounded by $1/3$.

The $D$ calls to the subroutine then take in total $D \cdot T(n, m, (3n\tau(M))^{-1})$ time. It takes in addition $D \cdot O(nm) = n^{1 + o(1)}$ operations to construct the vectors and $O(D^{\omega})$ time for step 2.\footnote{We always assume $f(x)$ can be exactly computed in constant time for any $x \in [0, 1]$.}
\end{proof}

We now combine the reduction above and the hardness result of BCP in Theorem \ref{thm:mainbcp}. The hardness of the KDE problem follows.
\begin{theorem} \label{thm:kdehardness}
Assuming SETH, for every $q > 0$, there exists $C \ge 0$ such that $\KDE_f(n, m, (3n\tau(M))^{-1})$
cannot be solved in time $O(n^{2-q})$ for any constant $q>0$. if either of the following holds: (1) $m > C \log n, D = m$ or (2) $m > C^{\log^* n}, D > m^{C^{\log^* n} \cdot (\log n)/m}$. Here $M$ is the $D \times D$ counting matrix in Definition \ref{def:countingmatrix}.
\end{theorem}

\section{Schur polynomials}
\label{sec:Schur}

Let $\F$ be a field. For matrices $A, B \in \F^{m \times n}$, let $A \circ B$ denote the Hadamard product of $A$ and $B$, i.e. the $m \times n$ matrix with entries $(A \circ B)[i, j] = A[i, j] \cdot B[i, j], i\in [m], j \in [n]$. For matrix $A \in \F^{m \times n}$ we define the Hadamard powers $A^{\circ 1} = A$ and $A^{\circ (k+1)} = A^{\circ k} \circ A$ for integer $k \ge 1$.
Similarly one can define the Hadamard product and Hadamard power for vectors $u \in \F^m$. Moreover, given a vector $u \in \F^m$ and a tuple $r = (r_1, \cdots, r_n) \in \N^{n}$,
we denote by $u^{\circ r}$ the $m \times n$ matrix
\[\begin{pmatrix}
| & | & & | \\
u^{\circ r_1} & u^{\circ r_2} & \cdots & u^{\circ r_n} \\
| & | & & |
\end{pmatrix}
=\begin{pmatrix}
u_1^{r_1} & u_1^{r_2} & \cdots & u_1^{r_n} \\
\vdots & \vdots & \ddots & \vdots \\
u_m^{r_1} & u_m^{r_2} & \cdots & u_m^{r_n}
\end{pmatrix}.\]
In particular, when $m = n$, we call $u^{\circ r}$ a generalized Vandermonde matrix. This is a natural generalization of the (usual) Vandermonde matrix $u^{\circ \delta}$ associated with the tuple $\delta = (0, 1, \cdots, m-1)$.

The concept of Schur polynomials was first proposed by Cauchy and defined as the ratio of two generalized Vandermonde determinants. In what follows we denote by $\N^m_< = \{x \in \N^m: x_1 < \cdots < x_m\}$ the set of $m$-tuples composed of distinct entries in ascending order, and define $\N^m_{\le} = \{x \in \N^m: x_1 \le \cdots \le x_m\}$ similarly.
\begin{definition}[Cauchy's definition of Schur polynomials]
Let $m>0$ be an integer and $\lambda = (\lambda_1, \cdots, \lambda_m) \in \N^m_{\le}$ an integer tuple.
We define the Schur polynomial on variables $(u_1, \cdots, u_m)$ by
\[s_{\lambda}(u) = s_{\lambda}(u_1, \cdots, u_m)
=\frac{\det(u^{\circ (\lambda + \delta)})}{\det(u^{\circ \delta})}
=\frac{\det(u^{\circ (\lambda + \delta)})}{V(u)}.\]
Here $V(u) = \det(u^{\circ \delta}) = \prod_{1 \le i < j \le m}(u_j - u_i)$ is the Vandermonde determinant.
\end{definition}

It is known that Schur polynomial has an equivalent definition by Littlewood using Young tableaux.
\begin{definition}
Let $m > 0$ be an integer, and $\lambda = (\lambda_1, \cdots, \lambda_m) \in \N^m_{\le}$ an integer tuple.
A semi-standard Young tableaux (SSYT) of shape $\lambda$ on alphabet $[m]$
is a left-aligned two-dimensional rectangular array $T$ of cells,
with $\lambda_i$ cells in the $i$-th row $(i \in [m])$, from bottom to top, such that
\begin{itemize}
\item each cell in $T$ is assigned with an entry from $1, \cdots, m$;
\item entries weakly decrease in each row, from left to right;
\item entries strictly decrease in each column, from top to bottom.
\end{itemize}
Moreover, for a SSYT $T$, we define the type $\mathrm t(T) = (t_1, \cdots, t_m) \in \N^m$ of $T$,
where $t_j$ is the number of cells in $T$ assigned with entry $j \in [m]$.
\end{definition}
\begin{definition}[Littlewood's definition of Schur polynomials]
Let $m > 0$ be an integer.
For integer tuple $\lambda = (\lambda_1, \cdots, \lambda_m) \in \N_{\le}^m$,
the Schur polynomial $s_{\lambda}(u)$ can be equivalently defined by
\[s_{\lambda}(u) = \sum_{T \in \mathcal T} u^{\mathrm t(T)},\]
where $\mathcal T$ is the set of all SSYT of shape $\lambda$ on alphabet $[m]$.
\end{definition}

\paragraph{Schur polynomial evaluated on special inputs}
On special inputs, the following closed form formulas for Schur polynomials are known.
\begin{proposition}
Let $m > 0$ be an integer and $\lambda \in \N^m_{\le}$ a tuple.
For any $q \in \F$ that has order at least $m$, we have
\[s_{\lambda}(1, q, q^2, \cdots, q^{m-1}) = \prod_{1 \le i < j \le m}
\frac{q^{\lambda_j + \delta_j} - q^{\lambda_i + \delta_i}}{q^{\delta_j} - q^{\delta_i}}. \tag{1}\]
For $q = 1$, we have
\[s_{\lambda}(\bo_N) = |\mathcal T| = \frac{V(\lambda + \delta)}{V(\delta)} \in \N. \tag{2}\]
Here $V(x) = \prod_{1\le i < j \le m} (x_j - x_i)$ is the Vandermonde determinant defined above.
\end{proposition}
Formula (1), (2) are respectively known as the Principal specialization formula
and the Weyl dimension formula. 


\begin{corollary}[first-order approximation \cite{KT21}]
Let $m > 0$ be an integer and $\lambda \in \N^m_{\le}$ an integer tuple.
For vector $u \in (\R_{\ge 0})^m_{\le} = \{x \in \R_{\ge 0}^m: x_1 \le \cdots \le x_m\}$, we have
\[u^{\lambda} \le s_{\lambda}(u) \le \frac{V(\lambda + \delta)}{V(\delta)} u^{\lambda}.\]
\end{corollary}
\begin{proof}
Among all SSYT of shape $\lambda$ on alphabet $[m]$, $u^{\mathrm{t}(T)}$ achieves its maximum when $T$ has each row filled with the largest number possible, whose type of $T$ is then exactly $\lambda$. The result then follows from $\max_{T \in \mathcal T} u^{\lambda} \le s_{\lambda}(u) \le |\mathcal T|\max_{T \in \mathcal T} u^{\lambda}$.
\end{proof}

\section{Direct calculation of $\tau(M)$: Gaussian kernel and $t$-Student kernel}
\subsection{Gaussian kernel}
In this section we show upper bounds for 
\[\tau(M) = \max_{b \in \R^D: b \ne 0} \frac{\|M^{-1}b\|_{\infty}}{\|b\|_{\infty}} \le D \cdot \max_{t, s \in [D]} |M^{-1}[t, s]|\]
where $M$ is the counting matrix associated with the Gaussian kernel $f(x) = e^{-Bx}$.
By standard facts in linear algebra, we can rewrite
\[|M^{-1}[t, s]| = \left|\frac{\det(M^{s-}_{t-})}{\det(M)}\right|\]
with $M_{t-}^{s-}$ the submatrix of $M$ consisting of all entries but those in row $s$ or column $t$.

For the simplicity of analysis, we study the counting matrix $M$ of dimension $(D+1)$. A slight modification of the main reduction can employ such a matrix to recover the distance matrix with a redundant row: $W[p, i] = \#[\|x^{(i)} - y^{(j)}\|_2^2 = p], i \in [n], p = \{0, 1, \cdots, D\}$.
\begin{theorem}
Let $\alpha \in \R^{D+1}$ be a fixed vector and $\beta \in \R^{D+1}$ be the identity vector.
Let $M$ be the $(D+1) \times (D+1)$ counting matrix associated with Gaussian kernel $f = e^{-Bx}$ and $\alpha, \beta$. Then there exists a vector $\alpha$ such that
\[\tau(M) \le \left(\frac{5e}{1 - e^{-B/D}}\right)^D.\]
\end{theorem}

\begin{proof}
Let $x_i = \exp(-B\alpha_i)$, then
$M = [x_i^j: i, j \in \{0, \cdots, D\}]$ is a Vandermonde matrix with
\[|\det(M)| = |\det(x^{\circ(0, \cdots, D)})| = V(x).\]
On the other hand, we observe that $M_{t-}^{s-}$ can be viewed as a generalized Vandermonde matrix. Making use of both the algebraic and combinatorial definition of Schur polynomials, we have\footnote{The Schur polynomial here in fact equals an elementary symmetric polynomial.}
\[|\det (M_{t-}^{s-})| = |\det((x^{s-})^{\circ(0, \cdots, t-1, t+1, \cdots, D)})|
= V(x^{s-}) \left( \sum_{R\subseteq \{0, \cdots, D\}\backslash\{s\}}\prod_{r\in R}x^r \right),\]
in which $x^{s-}$ denotes the vector $(x_0, \cdots, x_{s-1}, x_{s+1}, \cdots, x_D)$ and
\[\sum_{R\subseteq \{0, \cdots, D\}\backslash\{s\}}\prod_{r\in R}x^r 
\le \prod_{i \in \{0, \cdots, D\} \backslash \{s\}}(1 + x_i) \le 2^D.\]
Therefore
\[\max_{s, t} |\det (M_{t-}^{s-})|
\le 2^D \cdot \max_s V(x^{s-}) = 2^D \cdot V(x)
\left( \min_s \prod_{i \in [D]\backslash \{s\}} |x_i - x_s| \right)^{-1}. \]

We now pick $0 = \alpha_0 < \cdots < \alpha_{D-1} < \alpha_{D} = 1/D$ such that
$|x_{i+1} - x_i| = |\exp(-B\alpha_{i+1}) - \exp(-B\alpha_i)|
= \frac 1{D}(1 - e^{-B/D})$. Let $r = \frac 1{D}(1 - e^{-B/D})$, then
\begin{align*}
\prod_{i \in \{0, \cdots, D\} \backslash \{s\}} |x_i - x_s|
&= \left(\prod_{i=0}^{s-1} (s - i) r \right)
\left(\prod_{i=s+1}^{D} (i-s) r\right) \\
&= r^{D} \cdot s! (D-s)!
\ge r^{D} \cdot (\frac{D}{2e})^{D/2 \cdot 2} = (\frac{1 - e^{-B/D}}{2e})^{D}.
\end{align*}
Combining the calculations together, we have
\[\tau(M) \le D \cdot \max_{s, t}\left|\frac{\det(M^{s-}_{t-})}{\det(M)}\right| 
\le D \cdot 2^D\left( \min_s \prod_{i \in [D]\backslash \{s\}} |x_i - x_s| \right)^{-1} \le \left(\frac{5e}{1 - e^{-B/D}}\right)^D.\]
\end{proof}

The following hardness result for Gaussian KDE therefore arises from a combination of the general KDE hardness Theorem \ref{thm:kdehardness} and the bound on $\tau(M)$ above.
\begin{theorem}[Hardness of Gaussian KDE]
Let $f(x) = e^{-Bx}$ be the Gaussian kernel with $B \ge 1$.
Then assuming SETH, for every $q > 0$, there exists $C_1, C_1', C_2, C_3, C_3', C_4 \ge 0$ such that $\KDE_f(n, m, \eps)$ cannot be solved in time $O(n^{2-q})$ if either of the following holds:
\begin{enumerate}[(1)]
\item $m > C_1\log n, B < C_1'\log n, 1/\eps > (C_2m/B)^m$, or
\item $C_3\log^* n < m < C_3'(\log n)/(\log\log n)$, $\log\log(1/\eps) > (\log n)/m \cdot (\log m) \cdot C_4^{\log^*n}$.
\end{enumerate}
\end{theorem}
\begin{proof}
For regime (1), first pick appropriate $C_1'$ so that $B/m < 1$.
\[3n\tau(M) = 3n\left(\frac{5e}{1 - e^{-B/m}}\right)^m \le 3n\left(\frac{5e}{B/(em)}\right)^m
=3 \cdot 2^{C_1^{-1}m} \cdot \left(\frac {5e^2m}B\right)^m \le \left(\frac {C_2m}B\right)^m\]
for some constant $C_2 > 0$.
For regime (2), we have $B/D \le 1$ and thus
\[3n\left(\frac{5e}{1 - e^{-B/D}}\right)^D \le 3n\left(\frac{5e^2D}{B}\right)^D \le 3n(5e^2D)^D.\]
If $m < (\log n)/(\log \log n)$,
\[\log D = (\log n)/m \cdot (\log m) \cdot C^{\log^* m}
= (\log\log n)^2(1 - o(1)) \cdot C^{\log^* m} > \log \log n.\]
\[\log(3n\tau(M)) \le \log(3n) + D\log(5e^2D) < C'D\log(5e^2D)\]
For some $C' > 0$. Thus,
\[\log \log(3n\tau(M)) \le \log D + \log\log(5e^2D) \le (\log n)/m \cdot (\log m) \cdot C_4^{\log^* m}.\]

\end{proof}

\subsection{$t$-Student kernel}
For $t$-Student kernels $f(x) = 1/(1 + x^\rho)$, we prove a similar bound on $\tau(M)$. The key observation here is that both $M$ and $M_{s-}^{t-}$ are (scaled) Cauchy matrices. For vectors $a, b \in \R^n$, the Cauchy matrix associated with $a, b$ is defined by $M[i, j] = 1/(a_i + b_j), i, j \in [n]$. The following closed-form formula is known for its determinant.
\begin{theorem}[Cauchy determinant]
Let $n \ge 1$ be an integer. For $a, b \in \R^{n}$, let $M = [1/(a_i + b_j)]_{i, j \in [n]}$ be the associated Cauchy matrix. Then
\[\det(M) = \frac{\prod_{1 \le i < j \le n}(a_i - a_j)(b_i - b_j)}
{\prod_{i, j \in [n]}(a_i + b_j)}.\]
\end{theorem}
\begin{corollary}\label{cor:cauchyratio}
For $a, b \in \R^n$ with $a_i, b_j \in (0, 1), \forall i, j \in [n]$, 
and $s, t \in [n]$, we have
\begin{align*}\det\left[\frac1{1+a_ib_j}\right]_{i, j \in [n]}
&= \left(\prod_{i=1}^n \frac1{a_i}\right)\det\left[\frac1{a_i^{-1}+b_j}\right] \\
&= \left(\prod_{i=1}^n \frac1{a_i}\right)
\frac{\prod_{1 \le i < j \le n}(a_i^{-1} - a_j^{-1})(b_i - b_j)}
{\prod_{i, j \in [n]}(a_i^{-1} + b_j)} 
= \frac{\prod_{1 \le i < j \le n}(a_j - a_i)(b_i - b_j)}
{\prod_{i, j \in [n]}(1 + a_ib_j)}.
\end{align*}
\[\left|\frac{\det[(1+a_ib_j)^{-1}]_{i, j \in [n]}}
{\det[(1+a_ib_j)^{-1}]_{i \in [n]^{s-}, j \in [n]^{t-}}}\right|
= \left|\frac{\prod_{i\in [n]^{s-}}(a_i - a_s)\prod_{j\in [n]^{t-}}(b_j - b_t)}
{\prod_{i \in [n]}(1 + a_ib_t)\prod_{j \in [n]^{t-}}(1 + a_sb_j)}\right|.\]
Here $[n]^{s-} = [n] \backslash\{s\}, [n]^{t-} = [n] \backslash \{t\}$.
\end{corollary}
Before proving the bound on $\tau(M)$, we gather several simple inequalities to be used.
\begin{lemma} 
(I) Let $r \ge 1, a, b > 0$ be real numbers. Then $(a+b)^r \ge a^r + b^r$.
(II) Let $a, b \ge 0$ be integers. Then $a!b! \ge ((\lfloor\frac{a+b}2\rfloor)!)^2$. 
\end{lemma}
\begin{proof}
(I) Taking the derivative, one can show $f(x) = (1 + x)^r - x^r$ is increasing in $x$ for $x > 0$.
Thus $f(b/a) = (1 + b/a)^r - (b/a)^r \ge f(0) = 1$. The inequality follows.

(II) For $a = b = 0$, the inequality is trivial. If not, we have 
$\binom{a+b}{a} \le \binom{a+b}{\lceil(a+b)/2\rceil}$, $a!b! \ge 
(\lceil\frac{a+b}2\rceil)!(\lfloor\frac{a+b}2\rfloor)! \ge ((\lfloor\frac{a+b}2\rfloor)!)^2$.
\end{proof}
\begin{theorem}
Let $\rho \ge 1$ be a real number. Let $\alpha \in \R^D$ be a fixed vector and $\beta \in \R^D$ be the identity vector.
Let $M$ be the $D \times D$ counting matrix associated with $t$-Student kernel $f(x) = 1/(1+x^{\rho})$ and $\alpha, \beta$. Then there exists a vector $\alpha$ such that
\[\tau(M) \le (7e)^{2\rho D} .\]
\end{theorem}
\begin{proof}
The counting matrix is
\[M = [f(\alpha_i\beta_j)]_{i, j \in [D]} = 
\left[\frac{1}{1 + (\alpha_i\beta_j)^{\rho}}\right]_{i, j \in [D]}\]
For simplicity we assume $D$ is odd. Let $s, t \in [D]$. By corollary \ref{cor:cauchyratio}, we have
\[\left|\frac{\det(M)}{\det(M^{s-}_{t-})}\right|
=\left|\frac{\prod_{i\in [D]^{s-}}(\alpha_i^\rho - \alpha_s^\rho)
\prod_{j\in [D]^{t-}}(\beta_j^\rho - \beta_t^\rho)}
{\prod_{i \in [D]}(1 + \alpha_i^\rho\beta_t^\rho)
\prod_{j \in [D]^{t-}}(1 + \alpha_s^\rho \beta_j^\rho)}\right|.\]
We set $\alpha = \beta$ to be the scaled identity vector with $\alpha_i = \beta_i = i/D, \forall i \in [D]$.
(By rescaling $\alpha_i = i/D^2, \beta_i = i$ one can make $\beta$ the identity vector.)
Then
\[\prod_{i \in [D]}(1 + \alpha_i^\rho\beta_t^\rho)
\prod_{j \in [D]^{t-}}(1 + \alpha_s^\rho \beta_j^\rho) \le 2^{2D-1},\]
\begin{align*}\prod_{i\in [D]^{s-}}|\alpha_i^\rho - \alpha_s^\rho|
&=\prod_{i = 1}^{s-1}\frac{s^\rho - i^\rho}{D^{\rho}}
\prod_{i = s+1}^{D}\frac{i^\rho - s^\rho}{D^{\rho}}
\ge \prod_{i = 1}^{s-1}\frac{(s-i)^\rho}{D^{\rho}}
\prod_{i = s+1}^{D}\frac{(i-s)^\rho}{D^{\rho}} \\
&=\left(\frac{(s-1)!(D-s)!}{D^{D-1}}\right)^\rho
\ge \left(\frac{(\lfloor\frac{D-1}2\rfloor)!}{D^{D-1}}\right)^{\rho}
\ge \left(\frac{D-1}{2eD}\right)^{\rho(D-1)} \ge (3e)^{-\rho D}.
\end{align*}
Similarly,
\[\prod_{j\in [D]^{t-}}|\beta_j^\rho - \beta_t^\rho| \ge (3e)^{-\rho D}.\]
In conclusion, we have
\[\tau(M) \le D \cdot \max_{s, t}\left|\frac{\det(M^{s-}_{t-})}{\det(M)}\right|
\le D \cdot 2^{2D} \cdot (3e)^{2\rho D} \le (7e)^{2\rho D}.\]
\end{proof}

Combining the bound on $\tau(M)$ above and the general KDE hardness Theorem \ref{thm:kdehardness}, we obtain
\begin{theorem}[Hardness results of $t$-Student KDE]
Let $f(x) = 1/(1+x^\rho)$ be a $t$-Student kernel parameterized by $\rho \ge 1$ an absolute constant.
Then assuming SETH, for every $q > 0$, there exists $C_1, C_2, C_3, C_3', C_4 \ge 0$ such that $\KDE_f(n, m, \eps)$ cannot be solved in time $O(n^{2-q})$ if either of the following holds:
\begin{enumerate}[(1)]
\item $m > C_1\log n, 1/\eps > n^{C_2}$, or
\item $C_3\log^* n < m < C_3'(\log n)/(\log\log n)$, $\log\log(1/\eps) > (\log n)/m \cdot (\log m) \cdot C_4^{\log^*n}$.
\end{enumerate}
Here $C_1, C_3, C_3'$ are absolute constants while $C_2, C_4$ are dependent on $\rho$.
\end{theorem}
\begin{proof}
For regime (1), 
\[3n\tau(M) = 3n(7e)^{2\rho m} = 3n(7e)^{2\rho \cdot C_1\log n} = n^{C_2}\]
for some $C_2$ dependent on $\rho$. For regime (2), if $m < (\log n)/(\log \log n)$,
\[\log D = (\log n)/m \cdot (\log m) \cdot C^{\log^* m}
= (\log\log n)^2(1 - o(1)) \cdot C^{\log^* m} > \log \log n.\]
\[\log(3n\tau(M)) = \log(3n) + 2\log(7e)\rho D < C'_\rho D,\]
\[\log\log(3n\tau(M)) < (\log n)/m \cdot (\log m) \cdot C_4^{\log^* m}.\]
for some $C_4$ dependent on $\rho$.
\end{proof}

\section{Cauchy-Binet Expansion} \label{sec:cauchybinet}
To bound $\tau(M)$ for counting matrices that are associated with more general kernels, we establish the connection between the determinant $\det(M)$ and the Taylor coefficients of $f$ via Cauchy-Binet Expansion, based on a framework given in \cite{ACSS20}. Our improvement mainly derives from an observation that the terms in the expansion are in fact generalized Vandermonde determinants. The additional structure in the determinants allows for simplifications in analysis through Schur polynomials.

Define infinite matrices $A: [D] \times \N \rightarrow \R$,
$\tilde A: [D] \times \N \rightarrow \R$, and $B: \N \times [D] \rightarrow \R$ by
\[A[i, k] = \alpha_i^k, \quad
\tilde A[i, k] = \frac{f^{(k)}(c)}{k!}\alpha_i^k, \quad B[k, j] = \beta_j^k.\]
Suppose the Taylor series of $f(c + x)$ at $c \in [0, 1]$ is \emph{absolutely convergent} whenever $c + x \in [0, 1]$. Then for $i, j \in [D]$, assuming $c + \alpha_i \beta_j \in [0, 1]$, we have
\[M[i, j] = f(c + \alpha_i\beta_j)
= \sum_{k=0}^{\infty} \frac{f^{(k)}(c)}{k!}\alpha_i^k \beta_j^k
=\sum_{k=0}^{\infty}\tilde A[i, k]B[k, j].\]
By Cauchy-Binet formula, we expand $\det(M)$ as follows.
\begin{align}
\det(M) &= \sum_{\bn \in \N^D_<} \det \tilde A[[D]; \bn] \cdot \det B[\bn; [D]] \notag\\
&=\sum_{\bn \in \N^D_<} \left(\prod_{n\in \bn} \frac{f^{(n)}(c)}{n!}\right)
\det A[[D]; \bn] \cdot \det B[\bn; [D]]  \notag\\
&=\sum_{\bn \in \N^D_<} \left(\prod_{n\in \bn} \frac{f^{(n)}(c)}{n!}\right)
\det (\alpha^{\circ \bn})\det (\beta^{\circ \bn}) \notag\\
&=\sum_{\lambda \in \N^D_{\le}}
\left(\prod_{i=1}^D \frac{f^{(\lambda_i + \delta_i)}(c)}{(\lambda_i + \delta_i)!}\right)
\det (\alpha^{\circ (\lambda + \delta)})\det (\beta^{\circ (\lambda + \delta)})
\quad (\lambda = \bn - \delta) \notag\\
&=V(\alpha)V(\beta) \sum_{\lambda \in \N^D_{\le}}
\left(\prod_{i=1}^D \frac{f^{(\lambda_i + \delta_i)}(c)}{(\lambda_i + \delta_i)!}\right)
s_{\lambda}(\alpha)s_{\lambda}(\beta). \label{eq:CBexpansion1} 
\end{align}
The last step follows from the algebraic definition of Schur polynomials. Similarly,
\begin{equation}
\det(M_{t-}^{s-}) = V(\alpha^{s-})V(\beta^{t-}) \sum_{\lambda \in \N^{D-1}_{\le}}
\left(\prod_{i=1}^{D-1} \frac{f^{(\lambda_i + \delta_i)}(c)}{(\lambda_i + \delta_i)!}\right)
s_{\lambda}(\alpha^{s-})s_{\lambda}(\beta^{t-}). \label{eq:CBexpansion2}
\end{equation}

\subsection{Absolutely monotonic kernels}
At this point one may try bounding (\ref{eq:CBexpansion1}) and (\ref{eq:CBexpansion2}) using analytic properties of Schur polynomials. However, for a general function $f$, the arbitrary signs of Taylor coefficients largely complicate the analysis, making tight bounds out of reach.
Hence we first study the easy case where $\tau(M)$ is associated with an \emph{absolutely monotonic} kernel $f$, i.e., $f:[0, 1]\rightarrow [0, 1]$ satisfies $f^{(n)}(c) \ge 0$ for all $n \in \N$ and $c > 0$. We will shortly see that such kernels are in fact not artificial as all the positive definite kernels naturally reduce to this case.

We first use the following indentity of Schur polynomials to ``align'' (\ref{eq:CBexpansion1}) and (\ref{eq:CBexpansion2}).
\begin{proposition}
If $\lambda_1 \ne 0$, then
$s_{(\lambda_1, \cdots, \lambda_n)}(\alpha_1 = 0, \alpha_2, \cdots, \alpha_n) =0$.
If $\lambda_1 = 0$, then
$s_{(\lambda_1, \cdots, \lambda_n)}(\alpha_1 = 0, \alpha_2, \cdots, \alpha_n) =
s_{(\lambda_2, \cdots, \lambda_n)}(\alpha_2, \cdots, \alpha_n)$.
\end{proposition}
\begin{proof}
By the combinatorial definition of Schur polynomials, we have
\[s_{(\lambda_1, \cdots, \lambda_n)}(\alpha_1, \alpha_2, \cdots, \alpha_n)
=\sum_{T \in \mathcal T} \alpha^{\mathrm t(T)}\]
where $\mathcal T$ is the set of all SSYT of shape $\lambda = (\lambda_1, \cdots, \lambda_n)$
on alphabet $[n]$, and $\mathrm t(T)$ denotes the type of SSYT $T$.
Now set $\alpha_1 = 0$. If the letter 1 appears in a SSYT $T$, i.e. $\mathrm t(T)[1] > 0$, then the corresponding monomial vanishes when evaluated. Therefore the equation simplifies to
\[s_{(\lambda_1, \cdots, \lambda_n)}(\alpha_1, \alpha_2, \cdots, \alpha_n)
=\sum_{T\in \mathcal T_1} \alpha^{\mathrm t(T)}\]
where $\mathcal T_1$ is the set of all SSYT of shape $\lambda$ on alphabet $\{2, 3, \cdots, n\}$.

However, if $\lambda_1 > 0$, no sequence of length $n$ on alphabet $\{2, 3, \cdots, n\}$
satisfies the (strictly) decreasing constraint in the first column of the tableau.
In this case $s_{(\lambda_1, \cdots, \lambda_n)}(\alpha_1, \alpha_2, \cdots, \alpha_n) = 0$.
If $\lambda_1 = 0$, then $\mathcal T_1$ is essentially the set of all SSYT of shape $\lambda'
=(\lambda_2, \cdots, \lambda_n)$ over alphabet $\{2, 3, \cdots, n\}$.
By definition,
\[s_{(\lambda_1, \cdots, \lambda_n)}(\alpha_1, \alpha_2, \cdots, \alpha_n)
=\sum_{T\in \mathcal T_1} \alpha^{\mathrm t(T)}
=s_{(\lambda_2, \cdots, \lambda_n)}(\alpha_2, \cdots, \alpha_n).\]
\end{proof}
In consequence, if we fix $\alpha = (\alpha_1, \alpha_2, \cdots, \alpha_D)$
with $\alpha_1 = 0$, then
\begin{align*}
\det(M) &= V(\alpha)V(\beta) \sum_{\lambda \in \N^D_{\le}}
\left(\prod_{i=1}^D \frac{f^{(\lambda_i + \delta_i)}(c)}{(\lambda_i + \delta_i)!}\right)
s_{\lambda}(\alpha)s_{\lambda}(\beta)\\
&= V(\alpha)V(\beta) \sum_{\lambda^{1-} \in \N^{D-1}_{\le}}
\left(f(c) \prod_{i=2}^{D} \frac{f^{(\lambda_i + \delta_i)}(c)}{(\lambda_i + \delta_i)!}\right)
s_{\lambda^{1-}}(\alpha^{1-})s_{\lambda^{1-}}(\beta^{1-})\\
&= V(\alpha)V(\beta) f(c) \sum_{\lambda \in \N^{D-1}_{\le}}
\left(\prod_{i=1}^{D-1} \frac{f^{(\lambda_i + \delta_{i+1})}(c)}{(\lambda_i + \delta_{i+1})!}\right)
s_{\lambda}(\alpha^{1-})s_{\lambda}(\beta^{1-}).
\end{align*}
Here $\lambda^{1-}, \alpha^{1-}, \beta^{1-}$ are obtained by removing the first entry in the corresponding partition/vector $\lambda, \alpha, \beta$.
(In the last step we abuse the notation by renaming $\lambda^{1-}$ as $\lambda$.) Meanwhile,
\begin{align*}
\max_{s, t}\det(M_{t-}^{s-}) &\le \max_{s, t} V(\alpha^{s-})V(\beta^{t-}) \cdot
\max_{s, t} \sum_{\lambda \in \N^{D-1}_{\le}}
\left(\prod_{i=1}^{D-1} \frac{f^{(\lambda_i + \delta_i)}(c)}{(\lambda_i + \delta_i)!}\right)
s_{\lambda}(\alpha^{s-})s_{\lambda}(\beta^{t-}) \\
&= \left(\max_{s, t} V(\alpha^{s-})V(\beta^{t-})\right)
\sum_{\lambda \in \N^{D-1}_{\le}}
\left(\prod_{i=1}^{D-1} \frac{f^{(\lambda_i + \delta_i)}(c)}{(\lambda_i + \delta_i)!}\right)
s_{\lambda}(\alpha^{1-})s_{\lambda}(\beta^{1-}).
\end{align*}
In what follows, we let
\[E_\lambda = \left(\prod_{i=1}^{D-1} 
\frac{f^{(\lambda_i + \delta_{i+1})}(c)}{(\lambda_i + \delta_{i+1})!}\right)
s_{\lambda}(\alpha^{1-})s_{\lambda}(\beta^{1-}), ~~
F_\lambda = \left(\prod_{i=1}^{D-1} 
\frac{f^{(\lambda_i + \delta_i)}(c)}{(\lambda_i + \delta_i)!}\right)
s_{\lambda}(\alpha^{1-})s_{\lambda}(\beta^{1-})\]
denote the corresponding terms in the two sums. For absolutely monotonic function $f$, we have $F_\lambda > 0$ and $E_{\lambda} > 0$ for all $\lambda \in N^{D-1}_{\le}$. Thus,
\begin{equation}
\label{eq:detRatio}
\max_{s, t} \frac{\det(M^{s-}_{t-})}{\det(M)}
\le \frac 1{f(c)} \max_{s, t} \frac{V(\alpha^{s-})V(\beta^{t-})}{V(\alpha)V(\beta)}
\max_{\lambda \in \N^{D-1}_{\le}}\frac{F_\lambda}{E_\lambda}.
\end{equation}

\subsection{KDE hardness for positive definite kernels} \label{sec:pdkde}
To accommodate positive definite kernels, we slightly modify the reduction in Section \ref{sec:mainreduction}.
Let $k(x, y) = f(\|x-y\|_2^2)$ be a positive definite kernel. By Schoenberg's characterization (Theorem \ref{schoenberg}), we have $(-1)^k \cdot f^{(k)}(t) \ge 0$ for all $k \in \N, t \ge 0$. Then for the function $g(x) = f(1-x)$, it holds $g^{(k)}(t) = (-1)^k \cdot f^{(k)}(1-t) \ge 0$ for all $k \in \N, t \in [0, 1]$.
Namely $g$ is an absolutely monotonic kernel. 

Next we see how $\KDE_f$ can be related to $\tau(M_g)$. Let $M$ be the $D \times D$ counting matrix associated with $g$, and let $W \in \N^{D\times n}$ be a re-indexed distance count matrix defined by $W[p, i] = \#[j\in [n]: \|x^{(i)}-y^{(j)}\|_2^2 = D-p]$. Then the matrix product $U = M \times W$ gives 
\begin{align}
U[\ell, i] &= \sum_{p=1}^D M[\ell, p] \cdot W[p, i] \notag\\ 
&=\sum_{p=1}^D g\big(\alpha_\ell\beta_p\big) 
\sum_{j \in [n]} \ind\left[\|x^{(i)} - y^{(j)}\|_2^2 = D-p\right] \notag\\
&=\sum_{j \in [n]} g\left(c + \alpha_\ell \cdot \beta\left[D-\|x^{(i)} - y^{(j)}\|_2^2\right]\right).
\end{align}
If $\beta$ is the identity vector, then
\begin{align*}
U[\ell, i] &=\sum_{j \in [n]} g\left(c + \alpha_\ell \cdot (D-\|x^{(i)} - y^{(j)}\|_2^2)\right) \\
&=\sum_{j \in [n]} f\left(1 - c - \alpha_\ell D + \alpha_\ell \|x^{(i)} - y^{(j)}\|_2^2 \right).
\end{align*}
Constructing vectors $\tilde x_{\ell}^{(i)}, \tilde y_{\ell}^{(j)} \in \R^{m+1}$ defined by $\tilde x_{\ell}^{(i)}[k] = \sqrt{\alpha(\ell)} x^{(i)}[k]$ if $k \in [m]$ and $\tilde x_{\ell}^{(i)}[m+1] = \sqrt {1-c-\alpha_{\ell}D}$;
$\tilde y_{\ell}^{(j)}[k] = \sqrt{\alpha(\ell)} y^{(j)}[k]$ if $k \in [m]$ and $\tilde y_{\ell}^{(j)}[m+1] = 0$,
we have
\[U[\ell, i] = \sum_{j \in [n]}f(\|\tilde x_{\ell}^{(i)} - \tilde y_\ell^{(j)}\|_2^2).\]
In this way we show that the new product $U$ can also be computed using KDE subroutines, and the reduction proceeds the same way as in Section \ref{sec:mainreduction} to recover the (re-indexed) distance count matrix.
By similar analysis, we have the following hardness result relating the complexity of $\KDE_f$ to $\tau(M_g)$.
\begin{theorem} \label{thm:pdkdehardness}
Let $f: [0, 1]\rightarrow [0, 1]$ be a function and $g(x) = f(1-x)$.
Then assuming SETH, for every $q > 0$, there exists $C \ge 0$ such that $\KDE_f(n, m, (3n\tau(M))^{-1})$
cannot be solved in time $O(n^{2-q})$ for any constant $q>0$. if either of the following holds: (1) $m > C \log n, D = m$ or (2) $m > C^{\log^* n}, D > m^{C^{\log^* n} \cdot (\log n)/m}$. Here $M$ is the $D \times D$ counting matrix associated with function $g$.
\end{theorem}

\subsection{Hardness of Rational Quadratic kernel}
For $f, \alpha, \beta$ of certain forms, we give explicit bounds on the two terms in (\ref{eq:detRatio}). Regarding the ratio of Vandermonde determinants, we focus on $V(x^{s-})/V(x)$ for scaled identity $x$ defined by $x_p = p/D$.
\begin{proposition}
Let $\rho \ge 1$ be a real number and $x \in [0, 1]^D$ be the vector defined by $x_i = i/D$.
Then for $s \in [D]$ we have \[\frac{V(x^{s-})}{V(x)} = \prod_{i\in [D]\backslash \{s\}}|x_i - x_s|^{-1}
\le (3e)^{D}.\]
\end{proposition}
\begin{proof}
\begin{align*}
\prod_{i\in [D]\backslash \{s\}}|x_i - x_s| 
&=\prod_{i=1}^{s-1} \frac{s - i}{D}
\cdot \prod_{i=s+1}^D \frac{i - s}{D} \\
&= \frac{(s-1)!(D-s)!}{D^{D-1}}
\ge \frac{((\frac{D-1}2)!)^2}{D^{D-1}}
\ge \left(\frac{D-1}{2eD}\right)^{(D-1)}
\ge (3e)^{-D}.
\end{align*}
\end{proof}
Hence for vectors $\alpha$ defined by $\alpha_\ell = \ell/D, \ell\in [D]$ and $\beta$ defined by $\beta_p = p/D, p \in [D]$, we have
\[\max_{s, t} \frac{V(\alpha^{s-})V(\beta^{t-})}{V(\alpha)V(\beta)} \le (3e)^{2D}.\]

For the term involving Taylor coefficients, we fix a real number $\sigma \ge 1$ and focus on the absolutely monotonic function $f(x) = (2-x)^{-\sigma}$. By calculation, for $k \in \N$,
$f^{(k)}(x) = (2-x)^{-(\sigma + k)} \prod_{i=0}^{k-1}(\sigma + i)$.
Then for $c = 0$, we have
\[\frac{F_{\lambda}}{E_{\lambda}}
=\left(\prod_{i=1}^{D-1} 
\frac{f^{(\lambda_i + \delta_i)}(0)}{(\lambda_i + \delta_i)!}\right)
\bigg/\left(\prod_{i=1}^{D-1} 
\frac{f^{(\lambda_i + \delta_{i+1})}(0)}{(\lambda_i + \delta_{i+1})!}\right)
=\prod_{i=1}^{D-1}\left(2 \cdot \frac{1 + \lambda_i + \delta_i}{\sigma + \lambda_i + \delta_i}\right)
\le 2^{D-1}.\]
Combining the calculations, we obtain the following bound on $\tau(M)$.
\begin{theorem}
Let $\sigma \ge 1$ be a real number. 
Let $f:[0, 1] \rightarrow [0, 1]$ be the function $f(x) = (2-x)^{-\sigma}$, and let vectors $\alpha, \beta \in \R^D$ be defined by $\alpha_\ell = \ell/D, \beta_p = p/D$. Then the $D \times D$ counting matrix $M$ associated with $f, \alpha, \beta$ has
\[\tau(M) \le D \cdot 2^\sigma \cdot (3e)^{2D} \cdot 2^{D-1} \le 2^{\sigma}(7e)^{2D}.\]
\end{theorem}

Combining the bound on $\tau(M)$ associated with $f(x) = (1 + (1-x))^{-\sigma}$ with the hardness of positive definite KDE Theorem \ref{thm:pdkdehardness}, we obtain
\begin{theorem}[Hardness of Rational Quadratic KDE]
Let $f(x) = 1/(1+x)^\sigma$ be a Rational Quadratic kernel parameterized by $\sigma \ge 1$ an absolute constant.
Then assuming SETH, for every $q > 0$, there exists $C_1, C_2, C_3, C_3', C_4 \ge 0$ such that $\KDE_f(n, m, \eps)$ cannot be solved in time $O(n^{2-q})$ for any constant $q>0$. if either of the following holds:
\begin{enumerate}[(1)]
\item $m > C_1\log n, 1/\eps > n^{C_2}$, or
\item $C_3\log^* n < m < C_3'(\log n)/(\log\log n), \log\log(1/\eps) > (\log n)/m \cdot (\log m) \cdot C_4^{\log^*n}$.
\end{enumerate}
Here $C_1, C_3, C_3'$ are absolute constants while $C_2, C_4$ are dependent on $\sigma$.
\end{theorem}
\begin{proof}
For regime (1), 
\[3n\tau(M) = 3n\cdot 2^{\sigma}(7e)^{2m} = n^{C_2}\]
for some $C_2$ dependent on $\rho$. For regime (2), if $m < (\log n)/(\log \log n)$,
\[\log D = (\log n)/m \cdot (\log m) \cdot C^{\log^* m} > \log \log n.\]
\[\log(3n\tau(M)) = \log(3n) + \sigma + 2\log(7e)\rho D < C'_\sigma D,\]
\[\log\log(3n\tau(M)) < (\log n)/m \cdot (\log m) \cdot C_4^{\log^* m}.\]
for some $C_4$ dependent on $\rho$.
\end{proof}

\section*{Acknowledgement}
We would like to thank Amol Aggarwal for constructive discussions on Schur polynomials, and anonymous reviewers for helpful suggestions. This research was supported in part by a grant from the Simons Foundation (Grant Number 825870 JA) and a Google Research Scholar award.

\bibliographystyle{alpha}
\bibliography{ref}

\newpage
\appendix

\section{KDE lower bounds based on \cite{BIS17} approach}
\label{sec:BISreduction}
To facilitate the following discussion, we state here the decision version of Hamming Exact BCP and Hamming Approximate BCP. As there exists a straightforward binary search reduction between the optimization problem and the decision problem, the hardness results translate.
\begin{problem}[Hamming BCP, Decision Version]
$\emph{\textsf{Hamming-BCP-Dec}}(n, m, p)$:
Given two sets $A, B$ as in Problem \ref{prob:HBCP} as well as an integer $p \in [0, m]$, decide if there exists $i, j \in [n]$ such that $\|x^{(i)} - y^{(j)}\|_2^2 \le p$.
\end{problem}
\begin{problem}[Hamming Approximate BCP, Decision Version]
$\emph{\textsf{Hamming-Apx-BCP-Dec}}$$(n, m, \mu, p)$:
Given two sets $A, B$ and real number $\mu$ as in Problem \ref{prob:HABCP}, as well as an integer $p \in [0, m]$, 
distinguish two cases: (1) $\min_{i, j \in [n]} \|x^{(i)} - y^{(j)}\|_2^2 \le p$; (2) $\min_{i, j \in [n]} \|x^{(i)} - y^{(j)}\|_2^2 \ge (1 + \mu)p$.
\end{problem}
\begin{theorem}
Assuming SETH, for every $q \in (0, 1)$, there exist $C > 0, \mu > 0$ such that if $m > C\log n$, then both $\mathsf{Hamming}$-$\mathsf{BCP}$-$\mathsf{Dec}(n, m, p)$ and $\emph{\textsf{Hamming-Apx-BCP-Dec}}$$(n, m, \mu, p)$ require $n^{2-q}$ time.
\end{theorem}

The first hardness result for KDE is the following reduction from \textsf{Hamming-BCP-Dec} to $\KDE_f$ constructed in \cite{BIS17}. Given $A = \{x^{(1)}, \cdots, x^{(n)}\}, B = \{y^{(1)}, \cdots, y^{(n)}\} \in \{0, 1\}^m$ and integer $p \in [0, m]$, we denote $d_{\min} = \min_{i, j\in [n]} \|x^{(i)} - y^{(j)}\|_2^2$. Letting $K$ be the kernel matrix associated with $f$ and $\tilde x^{(i)} = m^{-1/2}x^{(i)}, \tilde y^{(j)} = m^{-1/2}y^{(j)}$, we have the following observation:
If $d_{\min} \le p$, then for fixed $i \in [n]$,
\[(K \times \bo)[i] = \sum_{j \in [n]} f(\|\tilde x^{(i)} - \tilde y^{(j)}\|_2^2)
= \sum_{j \in [n]} f\left(\frac 1m\|x^{(i)} - y^{(j)}\|_2^2\right)\ge f\left(\frac pm \right).\]
If $d_{\min} \ge p+1$, then
for fixed $i \in [n]$,
\[(K \times \bo)[i] = \sum_{j \in [n]} f\left(\frac 1m\|x^{(i)} - y^{(j)}\|_2^2\right)\le n \cdot f\left(\frac{p+1}{m}\right).\]

If $n \cdot f\left(\frac{p+1}{m}\right) < f\left(\frac{p}{m}\right)$, then one can all the KDE subroutine with parameters $n, m, \eps = \frac 1{3n} \left(f\left(\frac{p}{m}\right) - n \cdot f\left(\frac{p+1}{m}\right)\right)$ to distinguish between the two cases and give answer to the decision version of the BCP problem. In other words, we have the following theorem.
\begin{theorem} \label{thm:BISreduction}
Suppose $f$ satisfies $n \cdot f\left(\frac{p+1}{m}\right) < f\left(\frac{p}{m}\right)$ for all $p \in [0, m]$.
Then if $\KDE_f(n, m, \eps)$ can be solved in $T(n, m, \eps)$ time, then \emph{\textsf{Hamming-BCP-Dec}}$(n, m, p)$ can be solved in $T(n, m, \frac 1{3n} \left(f\left(\frac{p}{m}\right) - n \cdot f\left(\frac{p+1}{m}\right)\right))$ time.
\end{theorem}
In particular, for function $f(x) = e^{-Bx}$ with $B > 2m\log n$,
the condition $n \cdot f\left(\frac{p+1}{m}\right) < f\left(\frac{p}{m}\right)$ is satisfied. The bound on $\eps$ evaluates to
\[\left[\frac 1{3n} \left(f\left(\frac{p}{m}\right) - n \cdot f\left(\frac{p+1}{m}\right)\right)\right]^{-1}
= O(n) \cdot \exp\left(\frac{Bp}{m}\right)
\le O(n) \cdot e^B\]
\begin{theorem}
Assuming SETH, for every $q > 0$, there exists $C > 0$ such that if $m = C\log n, B = 2m\log n, 1/\eps = O(n) \cdot e^B$,
then $\KDE_f(n, m, \eps)$ cannot be solved in $O(n^{2-q})$ time.
\end{theorem}

Alman and Aggarwal \cite{AA22} make an improvement by combining the reduction of \cite{BIS17}
with the hardness of Hamming approximated BCP given by Rubinstein \cite{Rub18}. Formally, they consider the reduction from \textsf{Hamming-Apx-BCP-Dec} to $\KDE_f$ below.

If $d_{\min} \le p$, then for fixed $i \in [n]$,
\[(K \times \bo)[i] 
= \sum_{j \in [n]} f\left(\frac 1m\|x^{(i)} - y^{(j)}\|_2^2\right)\ge f\left(\frac pm \right).\]
If $d_{\min} \ge (1+\mu)p$, then
for fixed $i \in [n]$,
\[(K \times \bo)[i] = \sum_{j \in [n]} f\left(\frac 1m\|x^{(i)} - y^{(j)}\|_2^2\right)\le n \cdot f\left((1+\mu)\frac{p}{m}\right).\]
Based on this observation, we have the following algorithm.
\begin{theorem}
Suppose $\KDE_f(n,m,\eps)$ can be solved in $T(n, m, \eps)$ time. Then \emph{\textsf{Hamming-Apx-BCP-Dec}}$(n, m, \mu, p)$ can be solved in 
\begin{itemize}
\item either $n \cdot \binom{m}{\le p} = n \cdot \sum_{k=1}^p \binom{m}{k}$ time,
\item or $T(n, m, \frac 1{3n}\left(f\left(\frac pm\right)- n \cdot f\left((1 + \mu)\frac pm\right)\right))$ time.
\end{itemize}
Here the second item requires $f\left(\frac pm\right) > n \cdot f\left((1 + \mu)\frac pm\right)$.
\end{theorem}
\begin{proof}
The second item is based on the same algorithm as that in Theorem \ref{thm:BISreduction}. The first item can be achieved by the brute-force algorithm: store all vectors $y \in B$ in a lookup table, then for each $x \in A$ check each bucket $b$ with $\|x - b\|_2^2 \le p$ to see if it is empty.
\end{proof}
In particular, let $p_0 \in [0, p]$ be an integer. For the condition $f\left(\frac pm\right) > n \cdot f\left((1 + \mu)\frac pm\right)$ to hold for $p \ge p_0$, it requires $B > \frac{m \log n}{\mu p_0}$.
In this case, \textsf{Hamming-Apx-BCP-Dec}$(n, m, \mu, p)$ can be solved in time
\begin{align*}
&~~~~\max\left\{
\max_{p < p_0} n\cdot \binom{m}{\le p},
\max_{p \ge p_0} T(n, m, \frac 1{3n}\left(\exp\left(\frac pm\right)- n \cdot f\left((1 + \mu)\frac pm\right)\right))
\right\} \\
&\le \max\left\{nm \cdot \binom{m}{p_0}, T(n, m, O(ne^B)^{-1})\right\}.
\end{align*}
By picking appropriate $p_0$ so that $\binom m{p_0} < n^{0.5}$,
we can show the following hardness result for KDE.
\begin{theorem}
Let $p_0 = p_0(n, m)$ be the solution to $\binom{m}{p_0} = n^{0.5}$.
Assuming SETH, for every $q > 0$, there exists $C > 0, \mu > 0$ such that if $m = C\log n, B = \frac{m\log n}{\mu p_0}, 1/\eps = O(n) \cdot e^B$,
then $\KDE_f(n, m, \eps)$ cannot be solved in $O(n^{2-q})$ time.
\end{theorem}

As stated in Section \ref{sec:prevlb}, both reductions rely on a crucial condition that $f$ is rapidly decreasing, i.e. $f(\frac {p+1}m) / f(\frac pm) > n$ or $f((1+\mu)\frac pm)/f(\frac pm) > n$.
This condition turns out rather restrictive and precludes any result for smooth kernels, e.g. Rational Quadratic kernel and $t$-Student kernel, as well as small-scale Gaussian kernels with $B = o(\log n)$.

\section{hardness of low-dimensional BCP} \label{sec:WilliamsChen}
To establish lower bounds for low-dimensional KDE, we seek an analogue of Hamming BCP in low dimensions.
One of the major hardness results in dimension $d = o(\log n)$ concerns $\ell_2$-BCP.
\begin{problem}[$\ell_2$ (Exact) Bichromatic Closest Pair]
\label{prob:lBCP}
$\ell_2\emph{\textsf{-BCP}}(n, m, D)$: Given two sets $A = \{x^{(1)}, \cdots, x^{(n)}\}, B = \{y^{(1)}, \cdots, y^{(n)}\} \in \Z^m$ such that $\max_{i, j \in [n]} \|x^{(i)} - y^{(j)}\|_2^2 \le D$, compute $\min_{i, j \in [n]} \|x^{(i)} - y^{(j)}\|_2^2$.
\end{problem}
\begin{problem}[$\Z$ Orthogonal Vectors]
$\Z$\emph{\textsf{-OV}}$(n, m, E)$: Given two sets $A = \{x^{(1)}, \cdots, x^{(n)}\}, B = \{y^{(1)}, \cdots, y^{(n)}\} \in [-E, E]^m$, determine if there exist $i, j \in [n]$ such that $\ip{x^{(i)}}{y^{(j)}} = 0$.
\end{problem}

In this Section we show the hardness of $\ell_2$-BCP in low dimensions through the reductions below. (The \textsf{Hamming-OV} problem and the related Orthogonal Vectors Conjecture are stated in Section \ref{sec:hardproblems}.)
\[\textsf{Hamming-OV} \longrightarrow \textsf{\Z -OV} \longrightarrow \ell_2\textsf{-BCP}\]
This chain of reductions was first developed by Williams \cite{Wil18}, where the crucial step is to encode an $\mathsf{OV}$ instance by a collection of $\Z$-$\mathsf{OV}$ instances in lower dimensions. More concretely, \cite{Wil18} shows the following theorem.
\begin{theorem}[\cite{Wil18}]
Let $1 \le l \le m$. There is a reduction from \emph{\textsf{OV}}$(n, m)$ to $m^{O(m/l)}$ many instances of $\Z$\emph{\textsf{-OV}}$(n, l+1, m^{O(m/l)})$ in $n \cdot m^{O(m/l)}$ time.
\end{theorem}
Chen \cite{ChenMinIP} later improved this result by saving on the number of $\Z$\textsf{-OV} instances.
\begin{theorem}[\cite{ChenMinIP}]
Let $1 \le l \le m$. There is a reduction in
\[O\left(n \cdot l^{O(6^{\log^* m} \cdot m / l)} \cdot \poly(m)\right)\]
time from \emph{\textsf{OV}}$(n, m)$ to $l^{O(6^{\log^* m} \cdot m/l)}$ instances of $\Z$\emph{\textsf{-OV}}$(n, l+1, l^{O(6^{\log^* m} \cdot m / l)})$.
\end{theorem}
Combined with OVC, this reduction leads to the following hardness result for $\Z$\textsf{-OV}.
\begin{theorem}
Assuming OVC, for every $\delta > 0$, there is a $c \ge 1$ for which the following holds: if $m = c\log n$, $l \ge 7^{\log^* m}$ and $E = l^{O(6^{\log^* m} \cdot m/l)}$, then $\Z$\emph{\textsf{-OV}}$(n, l, E)$ cannot be solved in $n^{2-\delta}$ time.
\end{theorem}
Next, we have the following standard reduction from $\Z$\textsf{-OV} to $\ell_2$\textsf{-BCP}. 
\begin{theorem}
If $\ell_2$\emph{\textsf{-BCP}}$(n, m, D)$ can be solved in $T(n, m, D)$ time, then $\Z$\emph{\textsf{-OV}}$(n, m, E)$ can be solved in $m^4E^8 \cdot T(n, m^2, 4m^2E^4)$ time.
\end{theorem}
\begin{proof}
For each $x \in A, y \in B$, construct vector 
\[u_x = \big[x[1] \cdot x[1], \cdots, x[1]\cdot x[m], x[2] \cdot x[1], \cdots, x[m] \cdot x[m]\big] \in [-E^2, E^2]^{m^2}, \]
\[v_y = -\big[y[1] \cdot y[1], \cdots, y[1] \cdots y[m], y[2] \cdot y[1], \cdots, y[m] \cdot y[m]\big] \in [-E^2, E^2]^{m^2}.\]
Then
\[\ip{u_x}{v_y} = \sum_{k, \ell \in [m]}(x[k]x[\ell])(-y[k]y[\ell]) = -(\ip{x}{y})^2,\]
\[\|u_x - v_y\|_2^2 = \|u_x\|_2^2 + \|v_y\|_2^2 + 2(\ip{x}{y})^2.\]
Since $\|u_x\|_2^2, \|v_y\|_2^2 \le m^2E^4$, one can group the $u_x$ and $v_y$ vectors according to their norms, and find the minimum distance between each pair of vector sets
$U_s = \{u_x: \|u_x\|_2^2 = s\}, V_t = \{v_y: \|v_y\|_2^2 = t\}$.
There exists an orthogonal pair $x \in A, y \in B$ if and only if any of the minimum distances equals $s+t$.

This algorithm calls the subroutine for at most $m^4E^8$ pairs, with maximum distance $D = \max_{x \in A, y \in B} \|u_x - v_y\|_2^2$ bounded below $4m^2E^4$.
\end{proof}

Combining the two steps we can show the hardness of $\ell_2$\textsf{-BCP} assuming the Orthogonal Vectors Conjecture.
\begin{theorem}
Assuming OVC, for every $\delta > 0$, there is a $c \ge 1$ for which the following holds: if $m = c\log n$, $l \ge 49^{\log^* m}$ and $D = l^{O(6^{\log^* m} \cdot m/l)}$, then $\ell_2$\emph{\textsf{-BCP}}$(n, l, D)$ cannot be solved in $n^{2-\delta}$ time.
\end{theorem}
Simplifying the notation, we have an alternative statement.
\begin{theorem}
Assuming OVC, for every $\delta > 0$, there exist $C_1, C_2 \ge 1$ such that if $l \ge C_1^{\log^* n}$ and $D = l^{C_2^{\log^* n}\cdot (\log n)/l}$, then $\ell_2$\emph{\textsf{-BCP}}$(n, l, D)$ cannot be solved in $n^{2-\delta}$ time.
\end{theorem}

\section{Upper bound: Kernel Method}
\label{sec:polymethod}

\begin{definition}
Let $B \ge 1, \eps > 0$ be real numbers, and $f:[0, B] \rightarrow \R$ a real function.
We say a non-constant polynomial $p = p(x)$ (additively) $\eps$-approximates $f$ if
we have $|p(x) - f(x)| < \eps$ for all $x \in [0, B]$.
Based on this, the $\eps$-approximate degree of $f$, denoted by $d_{B;\eps}(f) \in \Z_{\ge 0}$, is defined to be the minimum degree of such a polynomial.
\end{definition}
\begin{remark}
To build an algorithm based on approximate polynomials, we also require such polynomials to be efficiently computable, i.e. all the coefficients can be computed in $\poly(d)$ time, where $d$ is the degree of the polynomial.
\end{remark}

\paragraph{Algorithm}
Suppose function $f:[0, B] \rightarrow \R$ has an $\eps$-approximate polynomial $p$
of degree $d$. Then for $u, v \in \R^m$,
\[p(\|u - v\|^2) = p\left(\sum_{k=1}^m (u[k] - v[k])^2\right)\]
is a polynomial of degree at most $2d$ in $2m$ variables.
By expanding this formula into a sum of monomials and grouping the entries in $u, v$ respectively in each term, we write
\[p(\|u - v\|^2) = \sum_{t \in T}c_t\phi_t(u)\psi_t(v)\]
where $T$ indexes all nonzero monomials inside, and $|T| \le \binom{2m+2d}{2m}$.
If we construct matrices $\Phi \in \R^{n \times |T|}, \Psi \in \R^{|T| \times n}$
defined by $\Phi[i, t] = c_t\phi_t(x^{(i)}), \Psi[t, j] = \psi_t(y^{(j)}), i, j \in [n], t\in T$,
then $P = \Phi \times \Psi$ is an approximation of $K$.

\paragraph{Running time}
First calculate all $c_t, t\in T$. For $p$ efficiently computable, this step can be completed in time $\poly(d) + O(|T|)$.
Next, evaluate all entries $\phi_t(x^{(i)}), \psi_t(y^{(j)}), t\in T, i, j \in [n]$.
For each entry we need to evaluate a monomial in $m$ variables using $O(m)$ time.
In total need $O(nm|T|)$ time.
Finally, compute $\Psi \times w \in \R^{|T|}$ and then $\Phi \times (\Psi \times w) = P \times w$.
Each takes $O(n|T|)$ time.
In total it requires $O(nm|T|) = n^{1+o(1)}|T|$ time if $d = n^{o(1)}$.

\paragraph{Approximation quality}
\begin{align*}
&~~~~\|K \times w - P \times w\|_{\infty} = \|(K - P) \times w\|_{\infty} \\
&\le \|w\|_1 \cdot \max_{i, j \in [n]} |f(\|x^{(i)} - y^{(j)}\|_2) - p(\|x^{(i)} - y^{(j)}\|_2)|
\le \eps\|w\|_1.
\end{align*}

\begin{theorem}
\label{thm:polymethod}
For function $f:[0, B] \rightarrow \R$, if $|T| = \binom{2m+2d_{B;\eps}(f)}{2m} < n^{o(1)}$,
then one can compute $\KDE_f(n, m, B, \eps)$ in $n^{1+o(1)}$ time. In particular, if $m = \Theta(\log n)$ and $d_{B;\eps} = o(m)$, then $|T| = \binom{2m+o(m)}{2m} < n^{o(1)}$ and $\KDE_f(n, m, B, \eps)$ can be computed in $n^{1+o(1)}$ time. 
\end{theorem}

We now combine the result with some known results regarding approximate degree. For Cauchy kernel, we have
\[d_{1; \eps}\left(\frac 1{1+x}\right) = \Theta(\log (1/\eps)).\]
as a folklore result. (A simple proof can be found in \cite[Sec. 5.2]{ACSS20}.)
For Gaussian kernel, \cite{AA22} shows
\[d_{B; \eps}(e^{-x}) = d_{1; \eps}(e^{-Bx}) = \Theta\left(\max\left\{
\sqrt{B\log (1/\eps)}, \frac{\log (1/\eps)}{\log(B^{-1}\log(1/\eps))}\right\}\right).\]
Hence by calculation we have the following upper bounds on running time.
\begin{theorem}
For the Cauchy kernel $f(x) = 1/(1+x)$, if $m = \Theta(\log n)$ and $\log(1/\eps) < 2^{o(m)}$, then $\KDE_f(n, m, B=1, \eps)$ can be computed in $n^{1+o(1)}$ time. For the Gaussian kernel $f(x) = e^{-x}$, if $m = \Theta(\log n)$ and $\log(1/\eps) < o(m)\log(m/B)$, then $\KDE_f(n, m, B, \eps)$ can be computed in $n^{1+o(1)}$ time.
\end{theorem}
\begin{proof}
For Cauchy kernel, the result is a straightforward corollary of Theorem \ref{thm:polymethod}.
For Gaussian kernel, it suffices to show that $d_{B;\eps}(e^{-x}) < o(m)$ when $1/\eps < o(m)\log(m/B)$.
Moreover, we wlog. focus on the case $\omega(m) < 1/\eps < o(m)\log(m/B)$ as $d_{B;\eps}$ is increasing in $1/\eps$.
Letting $r = m/B = \omega(1)$ and $s = (m\log r)/(1/\eps) = \omega(1)$, we have
\[\sqrt{B\log (1/\eps)} = \sqrt{\frac{m^2}{s} \cdot \frac{\log r}{r}} = \sqrt{o(m^2) \cdot o(1)} = o(m),\]
\[\frac{\log (1/\eps)}{\log(B^{-1}\log(1/\eps))}
=\frac{(m/s)\log r}{\log r + \log\log r - \log s} <\frac ms = o(m).\]
The last step follows from $s = (m\log r)/(1/\eps) < o(\log r)$.
\end{proof}

\end{document}